\documentstyle[epsf]{article}

\newcommand{\jmp}{J. Math. Phys.}

\newcommand{\pr}{Phys. Rev. D}

\newcommand{\newsection}{ \setcounter{equation}{0} \section}
\newcommand{\beq}{\begin{equation}} \newcommand{\eeq}{\end{equation}}
\newcommand{\bea}{\begin{eqnarray}} \newcommand{\eea}{\end{eqnarray}}

  \newcommand
{\Romannumeral}[1]{\uppercase\expandafter{\romannumeral#1}}

\newcommand{\be}{\begin{enumerate}} \newcommand{\ee}{\end{enumerate}}
\newcommand{\bi}{\begin{itemize}} \newcommand{\ei}{\end{itemize}}
\newcommand{\ba}{\begin{array}} \newcommand{\ea}{\end{array}}
\newcommand{\bc}{\begin{center}} \newcommand{\ec}{\end{center}}
\newcommand{\bt}{\begin{tabular}} \newcommand{\et}{\end{tabular}}

\def\lsim{\mathrel{\rlap{\lower4pt\hbox{\hskip1pt$\sim$}}
    \raise1pt\hbox{$<$}}}           
\def\gsim{\mathrel{\rlap{\lower4pt\hbox{\hskip1pt$\sim$}}
    \raise1pt\hbox{$>$}}}           

\newcommand{\Tr}{\mathop{\rm Tr}}           
\newcommand{\half}{\textstyle {1\over2} \displaystyle}    
\newcommand{\Dslash}{{\hbox{D}\kern-0.6em\raise0.15ex\hbox{/}}} 

\begin{document}

\setlength{\oddsidemargin}{0cm} \setlength{\baselineskip}{7mm}

\input epsf




\begin{normalsize}\begin{flushright}
    DAMTP-2004-61 \\
    UCI-2004-39 \\
    June 2004 \\
\end{flushright}\end{normalsize}

\begin{center}
  
\vspace{20pt}
  
{\Large \bf Non-Perturbative Gravity and the Spin of the Lattice Graviton }

\vspace{30pt}
  
{\sl Herbert W. Hamber}
$^{}$\footnote{e-mail address : hhamber@uci.edu} \\
Department of Physics and Astronomy \\
University of California \\
Irvine, CA 92697-4575, USA \\

and \\
{\sl Ruth M. Williams}
$^{}$\footnote{e-mail address : rmw7@damtp.cam.ac.uk} \\
Girton College, Cambridge CB3 0JG, and   \\
Department of Applied Mathematics and Theoretical Physics \\
Centre for Mathematical Sciences \\
Wilberforce Road, Cambridge CB3 0WA, United Kingdom.
\\

\end{center}

\vspace{15pt}

\begin{center} {\bf ABSTRACT } \end{center}
\vspace{12pt}
\noindent

The lattice formulation of quantum gravity provides a natural
framework in which non-perturbative properties of the
ground state can be studied in detail.
In this paper we investigate how the lattice results relate
to the continuum semiclassical expansion about smooth manifolds.
As an example we give an explicit form for the lattice ground state
wave functional for semiclassical geometries.
We then do a detailed comparison between the more recent predictions
from the lattice regularized theory, and
results obtained in the continuum for the non-trivial ultraviolet fixed
point of quantum gravity found using weak field and non-perturbative methods.
In particular we focus on the derivative of the beta function
at the fixed point and the related universal critical exponent $\nu$ for gravitation.
Based on recently available lattice and
continuum results we assess the evidence for the
presence of a massless spin two particle in the continuum
limit of the strongly coupled lattice theory.  
Finally we compare the lattice prediction for the vacuum-polarization
induced weak scale dependence of the gravitational coupling
with recent calculations in the continuum, finding similar effects.

\vspace{15pt}


\vfill

\pagestyle{empty}

\newpage

\pagestyle{plain}

\vskip 10pt
\newsection{Introduction}
\hspace*{\parindent}

It is widely believed that an understanding of the properties
of quantum gravitation would have important consequences
in many areas of cosmology and high energy physics.
Unfortunately approaches to quantum gravity based on linearized perturbation
methods have had moderate success so far,
as the underlying theory is known not to be perturbatively renormalizable
\cite{fey,hooft}.
A lack of perturbative renormalizability implies that an increasing
number of counterterms needs to be added in order to make the
theory finite order by order in perturbation theory. 
It has been recognized for some time though that the lack
of perturbative renormalizability is not necessarily
an obstacle in defining a consistent quantum theory \cite{wilson}, as several
simpler field theory models suggest \cite{parisi} (most notably the non-linear
sigma model above two dimensions) and  
recent rigorous results seem to support \cite{nonren}.
In the continuum non-perturbative renormalizability requires
the existence of a non-trivial ultraviolet fixed
point of the renormalization group.
In the presence of a lattice momentum cutoff, the corresponding
requirement is the existence of a phase transition
with a divergent correlation length. 

In the simplicial lattice formulation of quantum gravity one proceeds
in a way similar to ordinary lattice gauge theories, and introduces
a lattice ultraviolet regulator which in principle allows for controlled,
systematic analytical \cite{regge,rowi} and numerical \cite{hw84,lesh,critical}
non-perturbative calculations of ground state properties, anomalous
scaling dimensions and invariant correlations.
Once the specific lattice action has been chosen, numerically
exact results can be obtained on finite volume lattices which
then need to be judiciously extrapolated to the infinite
volume limit using the well-established methods of finite size scaling.
In this paper we address the basic issue of the relationship
between recent lattice results and a variety
of approximate perturbative and non-perturbative results
obtained in the continuum formulation, with
the intent of establishing a set of connections between
the two formulations that go beyond weak coupling 
perturbation theory and the perturbative expansion about smooth manifolds.

The starting point for a non-perturbative study of quantum gravity
is usually a suitable definition of the discrete Feynman path integral.
In the simplicial lattice approach one starts from the discretized
Euclidean path integral for pure gravity,
with the squared edge lengths taken as fundamental dynamical variables,
\beq
Z_L \; = \; \int_0^\infty \; \prod_s \; \left ( V_d (s) \right )^{\sigma} \;
\prod_{ ij } \, dl_{ij}^2 \; \Theta [l_{ij}^2]  \; 
\exp \left \{ 
- \sum_h \, \Bigl ( \lambda \, V_h - k \, \delta_h A_h 
+ a \, \delta_h^2 A_h^2 / V_h  + \cdots \Bigr ) \right \}  \;\; 
\label{eq:zlatt} 
\eeq
(see reference \cite{critical} for notation).
The above expression is supposed to represent a lattice discretization of the
continuum Euclidean path integral for pure quantum gravity
\beq
Z_C \; = \; \int \prod_x \;
\left ( {\textstyle \sqrt{g(x)} \displaystyle} \right )^{\sigma}
\; \prod_{ \mu \ge \nu } \, d g_{ \mu \nu } (x) \;
\exp \left \{ 
- \int d^4 x \, \sqrt g \, \Bigl ( \lambda - { k \over 2 } \, R
+ { a \over 4 } \, R_{\mu\nu\rho\sigma} R^{\mu\nu\rho\sigma}
+ \cdots \Bigr ) \right \}  \;\; ,
\label{eq:zcont}
\eeq
with $k^{-1} = 8 \pi G $, $G$ Newton's constant, and the $\cdots$
represent higher order curvature invariant terms.
The Regge lattice action only propagates spin two degrees of freedom
in the weak field limit, while the cosmological and measure
terms contain only local volume contributions.
The discrete gravitational measure in $Z_L$
can be considered as the lattice analog of the DeWitt \cite{dewittm,misnerm}
continuum functional measure \cite{hartle1,gauge,det}.
A cosmological constant term with bare $\lambda > 0$ is needed for
convergence of the path integral \cite{hw84,monte,monte1}.
Without loss of generality one can rescale the metric and set $\lambda=1$.
The curvature squared terms ($a \rightarrow 0$) allow one to control the
short distance fluctuations in the curvature, and as far as the
functional measure parameter $\sigma$ is concerned  most
of the recent work has focused on the case $\sigma=0$ 
(for more details on the choice of action and measure the reader is
referred to the review \cite{lesh}, and further references therein).

The present numerical evidence from the discrete model
of Eq.~(\ref{eq:zlatt}) suggests that quantum gravity
in four dimensions exhibits a phase transition in the coupling $G$
between two physically distinct phases \cite{critical}:
a strong coupling phase, in which the geometry becomes smooth at large
scales, 
\beq
\langle g_{\mu\nu} \rangle \; \approx \; c \; \eta_{\mu\nu}  \;\;  ( G > G_c )
\eeq
with a vanishingly small average curvature in the vicinity
of the critical point at $G_c$, and a weak coupling phase
\beq
\langle g_{\mu\nu} \rangle \; = \; 0 \;\;  ( G < G_c )
\eeq
in which the geometry becomes degenerate, bearing some resemblance
to a dilute branched polymer. 
It is clear that based on its geometric properties, only the smooth phase is
physically acceptable.
The existence of a phase transition at finite
coupling $G$ is usually associated with the appearance
of an ultraviolet fixed point of the renormalization group,
and implies in principle non-trivial scaling properties for
the coupling constant and invariant correlations.

In the lattice theory the presence of a fixed point or phase transition
is usually inferred (as in other lattice field theories)
from the appearance of non-analytic terms in
invariant local averages, such as the average scalar curvature
\beq
{ < \int d^4 x \, \sqrt{ g } \, R(x) >
\over < \int d^4 x \, \sqrt{ g } > }
\mathrel{\mathop\sim_{ k \rightarrow k_c}}
A_{\cal R} \; ( k_c - k ) ^{ 4 \nu - 1 } \;\; .
\label{eq:rk}
\eeq
Without such singularities the lattice continuum limit cannot be taken,
as one needs a divergent correlation length to define
the lattice continuum limit.
Indeed $k_c$ here is defined as the location of the non-analyticity in the
partition function and its averages, the latter often obtained
by differentiation with respect to a source or some other parameter.
A precise determination of $\nu$ then allows one
to connect singularities in averages such as the one above
to other long-distance properties of the theory.
In particular the relation between the critical exponent $\nu$
and the derivative of the renormalization group
beta function at the fixed point $ \beta ' (G_c) =  - 1/ \nu $
implies a scale dependence of Newton's constant (due to gravitational
vacuum polarization effects) of the form
\beq
G(r) \; = \; G(0) \left [ \, 1 \, + \, c \, ( r / \xi )^{1 / \nu} \, 
+ \, O (( r / \xi )^{2 / \nu} ) \, \right ] \;\; ,
\label{eq:grun}
\eeq
where $\xi$ is a renormalization group invariant scale parameter and
$c$ a calculable numerical constant of order one.
Detailed numerical studies of the Regge lattice gravity model give a value
very close to $\nu^{-1}=3$ \cite{critical}.
Since one finds for the critical value $ G_c \approx 0.626$ in units of the
lattice spacing, one would conclude that the lattice
theory is not weakly coupled in the vicinity of the fixed point.
It seems natural to interpret the momentum scale $\xi^{-1}$ as
arising due to a gravitational analogue of dimensional transmutation, and
it playing a role in gravitation similar to the universal
scaling violation parameter $\Lambda_{\overline{MS}}$
of QCD \cite{frampton}.
Other lattice approaches to quantum gravity based on discrete dynamical
triangulations with fixed edge lengths and which give
rise to a rather different phase structure are reviewed in \cite{dtrs}.

In this paper we examine a number of fundamental issues that have a bearing
on the relationship between lattice and continuum models for quantum 
gravity. 
First we will consider the lattice analog of the semiclassical expansion
for the ground state wave functional of continuum gravity.
Within the continuum formulation, the semiclassical expansion about
smooth manifolds with bounded quantum fluctuations 
allows one to exhibit in a clear and direct way the transverse
traceless modes (or equivalently spin two modes)
as the only physical gravitational degrees of freedom.
Such an expansion is most easily carried out with the Euclidean
functional integral approach, wherein the gravitational
action is expanded in the weak field metric, and the
resulting Gaussian integrals are subsequently carried out.

In trying to construct the lattice analog of the ground state
functional for semiclassical gravity one has two options.
The first procedure relies on constructing directly a
lattice expression for the exponent of the ground state functional,
obtained by transcribing the continuum expression
in terms of lattice variables. 
The unique procedure we follow here is to proceed from
the lattice expression for the gravitational action, specified
on a fixed time slice, and supplemented by the appropriate vacuum
gauge conditions.
A crucial ingredient in this method is the correct identification
of the correspondence between continuum degrees of freedom
(the metric) and the lattice variables (the edge lengths).
This correspondence is fixed by the relationship between
the lattice Regge action and the continuum Einstein action,
at least in the weak field limit.
The resulting lattice expression is then equivalent to the continuum
one by construction.

The second procedure relies instead only on the expression for the
lattice gravitational action, as computed in the weak field limit,
and determines the explicit lattice
form for the ground state functional for linearized gravity
by performing explicitly the necessary lattice Gaussian functional integrals.
The resulting discrete expression can then be compared to
the continuum one by re-expressing the edge lengths in terms
of the metric. 
It is encouraging that the resulting lattice expression completely agrees
with what is found by using the previous method.

It is advantageous in performing the above calculation to
introduce spin projection operators, which separate out
the spin zero, spin one and spin two components of the
gravitational action.
As a by-product one can then show that the lattice gravitational action 
only propagates massless spin two (or transverse-traceless)
degrees of freedom in the weak field limit, as is the case in the continuum. 
Furthermore, as expected, the lattice ground state functional
for linearized gravity only contains those physical modes. 

In subsequent sections of the paper we examine systematically
the relationship
between recent non-perturbative results obtained in the lattice theory
and corresponding calculations performed in the continuum theory.
The latter suggest the presence of a non-trivial ultraviolet
fixed point in $G$, and in some cases have even led to definite
predictions for the universal critical exponent of quantum
gravitation, which can therefore be compared quantitatively
to the lattice results.

Besides relying on the recent lattice and continuum results
for quantum gravitation, one can also try to independently
estimate the gravitational scaling dimensions using 
what is known based on exact and approximate
renormalization group methods for spin zero (self-interacting scalar field
in four dimensions) and spin one 
(Abelian non-compact gauge theories), for which
a wealth of information is available on the critical indices.
Based on this comparison, we will argue that these results too are consistent
with what is known about the gravitational exponents in four dimensions.
Finally we describe a simple geometric argument
which interprets the value found for the gravitational exponent $\nu^{-1}=3$.

\vskip 30pt
\newsection{Ground State Wave Functional of Linearized Gravity}
\hspace*{\parindent}

According to the path integral prescription for Euclidean quantum gravity,
the wave function of the
state of minimum excitation for an asymptotically flat three-geometry
with specified metric is 
\beq
\Psi_0 [ {}^3 g_{ij}, t] \; = \; {\cal N} \; \int_{ {}^3 g_{ij} }
 [ d g_{\mu\nu} ] \; e^{- I[g_{\mu\nu}]}
\eeq
where the integral is over all Euclidean four-geometries which are
asymptotically flat and bounded at time $t$ by an asymptotically flat
hypersurface with induced metric $ {}^3 g_{ij} $.

Kucha\u r \cite{kuchar} has given an expression for the ground-state wave
functional of linearized gravity.
In the vacuum or ``Coulomb'' gauge
$ h_{ik,k}=0 $, $ h_{ii}=0 $, $ h_{0i}=0 $ the ground
state functional is given by 
\beq
\Psi_0 [ h_{ij}^{TT}, t] 
\; = \; {\cal N} \; \exp 
\left \{ - { 1 \over 4 l_P^2 }
\int d^3 {\bf k} \; \omega_{\bf k} \; 
h_{ij}^{TT} ({\bf k}) \; \bar h^{TT}_{ij} ({\bf k}) \right \}
\label{eq:kuchar}
\eeq
where $h_{ij}^{TT} ({\bf k})$ is a Fourier component of the transverse
traceless part of the deviation of the three-metric from the flat three-metric
in rectangular coordinates,
\beq
h_{ij} ({\bf x} , t) \; = \; {}^3 g_{ij} ({\bf x} , t) - \delta_{ij} \; ,
\eeq
$\omega_{\bf k}= | {\bf k} | $, ${\cal N}$ is a normalization factor, and 
$l_P = ( 16 \pi G )^{1/2} $ is the Planck length in a system of
units where $\hbar = c =1$.
Equivalently one can write, in real space and in terms of first derivatives
of the fields, the expression
\beq
\Psi_0 [ h_{ij}^{TT}, t] \; = \;
{\cal N} \exp \left ( - {1 \over 8 \pi^2 l_P^2 }
\int d^3 {\bf x} \int d^3 {\bf x}'
\; { h^{TT}_{ij,k} ( {\bf x} ) h^{TT}_{ij,k} ({\bf x}') 
\over | {\bf x} - {\bf x'} |^2 } \right )  \;\; .
\label{eq:kuchar1}
\eeq

\noindent
The above ground state wave function of linearized gravity was originally
evaluated by Kucha\u r using canonical methods \cite{kuchar}. Later
the same formula was obtained by Hartle using the Euclidean path
integral prescription \cite{hartle} (see Section 3).

\vskip 30pt
\subsection*{Electromagnetic Case}
\hspace*{\parindent}

It is instructive, in view of the calculations to be done for the
gravitational case
in the next sections, to consider as an aside the much simpler case of
electromagnetism. 
Indeed a completely analogous set of results for the ground state functional
holds for the electromagnetic case, and further
brings out the relationship between the action and the quantity
appearing in the exponent of the wave functional, which
can be described as the time-slice action contribution.

In the Coulomb gauge $ \partial_i A_i = 0 $, $ A_0=0$ the ground
state functional is given by Kucha\u r in terms of the transverse parts of the potentials only
\beq
\Psi_0 [ {\bf A},t ] \; = \; {\cal N} \exp \left ( - {1 \over 2} \int d^3 {\bf k} 
\; k \; A^T_i ( {\bf k} ) \bar A^{T}_i ({\bf k}) \right )
\label{eq:kuchar_em}
\eeq
or, equivalently, in terms of the $\bf B$ fields as
\beq
\Psi_0 [ {\bf B},t ] \; = \; {\cal N} \exp \left ( - {1 \over 4 \pi^2 }
\int d^3 {\bf x} \int d^3 {\bf x}'
\; { {\bf B} ( {\bf x} ) \cdot {\bf B} ( {\bf x}') \over | {\bf x} - {\bf x}' |^2 }
\right ) \;\; .
\eeq
It is easy to see that the expression in the exponent of
Eq.~(\ref{eq:kuchar_em}) is related in a simple
way to the original electromagnetic action. 
One has for the action appearing in the exponent of the Feynman path integral
\beq
I[A_{\mu}] \; = \; { 1 \over 4} \int d^4 x \; F_{\mu\nu} (x) \; F^{\mu\nu} (x)
\eeq
which for a single mode reduces to
\beq
I_k \; = \; { 1 \over 4} \; 
\bigl ( k_\mu A_\nu (k) - k_\nu A_\mu (k) \bigr )^2  \;\; .
\eeq
In terms of the transverse projection of the field
\beq
A^T_i \; = \; ( \delta_{ij} - { k_i k_j \over {\bf k}^2 } ) \;  A_j
\eeq
one has for $k_0 = A_0 =0$ 
\beq
{\bf k}^2 A^T_i ( {\bf k} ) A^T_i ( {\bf k} ) 
\; = \; { 1 \over 2} \; \bigl ( k_i A_j ( {\bf k} ) - k_j A_i ( {\bf k} )
\bigr )^2
\eeq
and therefore, after summing over all modes,
\beq
\int d^3 {\bf k} \; \omega_{\bf k} \; A^T_i ( {\bf k} ) \bar A^T_i ( {\bf k} )
\; = \;  \int { d^3 {\bf k}  \over 2 \omega_{\bf k}} 
 \; \bigl ( k_i A_j ( {\bf k} ) - k_j A_i ( {\bf k} ) \bigr )^2
\label{eq:slice_em}
\eeq
with $\omega_{\bf k} = \sqrt{{\bf k}^2}$. 
Thus the expression appearing in the exponent of 
Eq.~(\ref{eq:kuchar_em}) is the same as the time-slice
contribution derived in Eq.~(\ref{eq:slice_em}).

\vskip 30pt
\subsection*{Spin Projections}
\hspace*{\parindent}

Returning to the case of the gravitational field one
can follow the same procedure, first in the continuum
and then on the lattice, and derive a lattice expression
for the vacuum functional for linearized gravity.
In the continuum the ground state functional
for linearized gravity Eq.~(\ref{eq:kuchar}) can be obtained
from the continuum action by suitably expanding the gravitational
action in the weak field limit, and then imposing the
appropriate gauge conditions. 
By later following the same procedure on the lattice, the
corresponding discrete expression can be derived.

The first step involves therefore the expansion of the
continuum Lagrangian $ - \sqrt{g} R $ in the weak field limit
\beq
g_{\mu\nu} \; = \; \eta_{\mu\nu} + \kappa h_{\mu\nu}
\eeq
with $\kappa = \sqrt{8 \pi G} $ and $| h_{\mu\nu} | $ small.
The quadratic part \cite{hooft,veltman} is given by
\beq
{\cal L}_{sym} \; = \;
 - {1 \over 2 } \; \partial_\mu h_{\mu\nu} \partial_\nu h_{\rho\rho}
\; + \; { 1 \over 2 } \; \partial_\mu h_{\mu\nu} \partial_\rho h_{\rho\nu}
\; - \; { 1 \over 4 } \; \partial_\mu h_{\nu\rho} \partial_\mu h_{\nu \rho}
\; + \; { 1 \over 4 } \; \partial_\mu h_{\nu\nu} \partial_\mu h_{\rho \rho}
\label{eq:wfe}
\eeq
with residual gauge invariance
\beq
h_{\mu\nu} \; \rightarrow \; h_{\mu\nu} \; + \; 
\partial_\mu \xi_\nu + \partial_\nu \xi_\mu
\eeq
and $\xi_\nu$ an arbitrary gauge function.
For one mode with wave-vector $k$ one has
\beq
{\cal L}_{sym} \; = \; {1 \over 2 } \; k_\mu h_{\mu\nu} k_\nu h_{\rho\rho}
\; - \; { 1 \over 2 } \; k_\mu h_{\mu\nu} k_\rho h_{\rho\nu}
\; + \; { 1 \over 4 } \; k_\mu h_{\nu\rho} k_\mu h_{\nu \rho}
\; - \; { 1 \over 4 } \; k_\mu h_{\nu\nu} k_\mu h_{\rho \rho}  
\label{eq:wfe1}
\eeq
\beq
\; = \; { 1 \over 2 } \; k_1^2 \;
( - h_{23}^2 - h_{24}^2 - h_{34}^2 + h_{22} h_{33}
+ h_{22} h_{44} + h_{33} h_{44} ) + \cdots  \;\;\;  .
\eeq
A gauge fixing term can be added of the form
\beq
{\cal L}_{fix} \; = \;
- { 1 \over 2 } \; ( k_\nu h_{\mu\nu} - {1 \over 2} k_\mu h_{\nu\nu} ) \;
( k_\rho h_{\mu\rho} - {1 \over 2} k_\mu h_{\rho\rho} )
\eeq
giving for the combined gauge-fixed weak field action
\beq
{\cal L}_{tot} \; = \; {\cal L}_{sym} \; + \; {\cal L}_{fix} \; = \; 
- {1 \over 2 } \; \partial_\mu h_{\alpha\beta} \partial_\mu h_{\alpha\beta}
\; + \; 
{ 1 \over 8 } \; \partial_\mu h_{\alpha\alpha} \partial_\mu h_{\beta\beta} \;\; .
\eeq
However in the following we shall rely instead on the vacuum
(or Coulomb) gauge fixing which
gives the ground state functional for linearized gravity discussed
previously.

It will be advantageous to define three independent spin projection
operators, which show explicitly the unique decomposition of
the continuum gravitational action for linearized gravity into spin two
(transverse-traceless) and spin zero (conformal mode) parts \cite{vnh}.
The spin-two projection operator is defined as
\bea
P^{(2)}_{\mu\nu\alpha\beta} \; = \; {1 \over 3 k^2 } 
\left ( k_\mu k_\nu \delta_{\alpha\beta} \; + \;
k_\alpha k_\beta \delta_{\mu\nu} \right ) \; - \;
{1 \over 2 k^2 } \left ( k_\mu k_\alpha \delta_{\nu\beta}
\; + \; k_\mu k_\beta \delta_{\nu\alpha}
\; + \; k_\nu k_\alpha \delta_{\mu\beta}
\; + \; k_\nu k_\beta \delta_{\mu\alpha} \right )  
\nonumber \\
\; + \; {2 \over 3 k^4 } k_\mu k_\nu k_\alpha k_\beta
\; + \; {1 \over 2} \left ( \delta_{\mu\alpha} \delta_{\nu\beta}
\; + \; \delta_{\mu\beta} \delta_{\nu\alpha} \right )
\; - \; {1 \over 3 } \delta_{\mu\nu} \delta_{\alpha\beta}  \;\; ,
\label{eq:s2}
\eea
the spin-one projection operator as
\beq
P^{(1)}_{\mu\nu\alpha\beta} \; = \; {1 \over 2 k^2 } \left (
k_\mu k_\alpha \delta_{\nu\beta} \; + \;
k_\mu k_\beta  \delta_{\nu\alpha} \; + \; 
k_\nu k_\alpha \delta_{\mu\beta} \; + \;
k_\nu k_\beta  \delta_{\mu\alpha} \right )
\; - \; {1 \over k^4 } k_\mu k_\nu k_\alpha k_\beta
\label{eq:s1}
\eeq
and the spin-zero projection operator as
\beq
P^{(0)}_{\mu\nu\alpha\beta} \; = \; - {1 \over 3 k^2 } 
\left ( k_\mu k_\nu \delta_{\alpha\beta} \; + \;
k_\alpha k_\beta \delta_{\mu\nu} \right )
\; + \; {1 \over 3 } \delta_{\mu\nu} \delta_{\alpha\beta} 
\; + \; {1 \over 3 k^4 } k_\mu k_\nu k_\alpha k_\beta  \;\; .
\label{eq:s0}
\eeq
The sum of the three spin projection operators is then equal to unity
\beq
P^{(2)}_{\mu\nu\alpha\beta} \; + \; 
P^{(1)}_{\mu\nu\alpha\beta} \; + \; 
P^{(0)}_{\mu\nu\alpha\beta} \; = \; 
{1 \over 2} \left ( \delta_{\mu\alpha} \delta_{\nu\beta}
\; + \; \delta_{\mu\beta} \delta_{\nu\alpha} \right )  \;\; .
\eeq
As a result one can define for the metric three orthogonal
fields of definite spin, the transverse-traceless (spin two) part
\beq
h^{TT}_{\mu\nu} \; = \;  P_{\mu\alpha} h_{\alpha\beta} P_{\beta\nu} -
{1 \over 3 } P_{\mu\nu} P_{\alpha\beta} h_{\beta\alpha}
\eeq
the longitudinal part (spin one)
\beq
h^{L}_{\mu\nu} \; = \; h_{\mu\nu} - P_{\mu\alpha} h_{\alpha\beta} h_{\beta\nu}
\eeq
and the trace (spin zero) part
\beq
h^{T}_{\mu\nu} \; = \; {1 \over 3 } P_{\mu\nu} P_{\alpha\beta} h_{\alpha\beta}
\eeq
such that their sum gives $h$
\beq
h \; = \; h^{TT} + h^{L} + h^{T}  \;\; .
\eeq
Here we have defined the projection operator
\beq
P_{\mu\nu} \; = \; \delta_{\mu\nu} - { 1 \over \Delta } \; 
{ \partial_\mu \partial_\nu }
\label{eq:proj}
\eeq
or equivalently in momentum space
\beq
P_{\mu\nu} \; = \; \delta_{\mu\nu} - { k_\mu k_\nu \over k^2 }
\label{eq:proj1}
\eeq
Using the three spin projection operators, the action for linearized
gravity can then be re-written simply as
\beq
{\cal L}_{sym} \; = \; - { 1 \over 4}
h_{\mu\nu} \left [ P^{(2)} - 2 P^{(0)} \right ]_{\mu\nu\alpha\beta} k^2
\;  h_{\alpha\beta}
\label{eq:wfe2}
\eeq
Imposing the gauge condition $h_{i 0}= h_{00}= h_{ik,k}=0$ and setting
$k_0=0$ one obtains
\beq
- { 1 \over 4} k^2 \; h^{TT}_{ij} h^{TT}_{ij}
+ { 1 \over 2} k^2 \; h^{T}_{ij} h^{T}_{ij}  
\eeq
with the second (spin zero) vanishing after further imposing the
trace condition $h_{ii} =0 $.
The resulting expression is then identical, up to a factor,
to the expression appearing in the exponent of the ground
state functional of linearized gravity of Eq.~(\ref{eq:kuchar}).

\vskip 30pt
\subsection*{Lattice Transverse Traceless Modes}
\hspace*{\parindent}

In this section the lattice analog of the TT-mode action will be derived
from the Regge lattice gravitational action, and a lattice expression for the 
gravitational wave functional will be given.
As a first step one needs to perform the weak field expansion for
the Regge action
\beq
I_R = \sum_{h} A_{h} \delta_{h}
\eeq
where $A_{h}$ is the area of the hinge $h$, and
$\delta_{h}$ is the deficit angle at the same hinge.
Following \cite{rowi} each hypercube
in a hypercubic lattice is divided up into
24 four-simplices, with vertices at $(0,0,0,0)$, $(0,0,0,1)$ ... $(1,1,1,1)$
(without loss of generality one can take the lattice spacing to be one).
The lengths of the 15 edges connecting the vertices $i$ and $j$ are denoted by
$l_{ij}$, where $i$ and $j$ range from 1 to 15, with the coordinates of
the endpoints interpreted as binary numbers (for more details, see Section 3).
Next each link length is allowed to fluctuate by an amount
$ 1 + e $ around the hypercubic lattice value. 
To lowest order in the edge fluctuation, the lattice action is
given by a quadratic form
\beq
I_R = {1 \over 2 } \; \sum_{ij} e_i \; M_{ij} \; e_j
\eeq
with $M$ a local matrix connecting only nearest-neighbor points.
In Fourier space one can write for
each of the fifteen displacements $e_i^{a,b,c,d}$,
defined at the vertex of the cube with labels $(a,b,c,d)$,
\beq
e_i^{(a,b,c,d)} = 
( \omega_1 )^a ( \omega_2 )^b ( \omega_4 )^c ( \omega_8 )^d e_i^0
\eeq
with $\omega_1 = e^{i k_1} $, $\omega_2 = e^{i k_2} $,
$\omega_4 = e^{i k_3} $, $\omega_8 = e^{i k_4} $
(we use the binary notation for $\omega$ and $e$,
but the regular notation for $k_i$.
For one mode (one set of $\omega$'s) one obtains therefore 
(see Appendix B in \cite{rowi})
\bea
6 e_1^2 + 16 e_3^2 + 18 e_7^2 + (
\omega_1 \omega_4 + \omega_2 \omega_4 +
\omega_1 \omega_8 + \omega_2 \omega_8 +
\bar \omega_1 \bar \omega_4 + \bar \omega_2 \bar \omega_4 +
\bar \omega_1 \bar \omega_8 + \bar \omega_2 \bar \omega_8 ) e_1 e_2
\nonumber \\
- ( 8 + 4 \omega_2 + 4 \bar \omega_2 ) e_1 e_3
- ( 2 \omega_1 + 2 \bar \omega_1 + 2 \omega_2 \omega_4 +
2 \bar \omega_2 \bar \omega_4 ) e_1 e_6 - 
( 12 + 6 \omega_4 + 6 \bar \omega_4 ) e_3 e_7 + \cdots
\eea
Each coefficient is real, as expected from the reality of the action.
Thus, for example in the above expression, we have
\bea
& \omega_1 \omega_4 + \omega_2 \omega_4 +
\omega_1 \omega_8 + \omega_2 \omega_8 +
\bar \omega_1 \bar \omega_4 + \bar \omega_2 \bar \omega_4 +
\bar \omega_1 \bar \omega_8 + \bar \omega_2 \bar \omega_8 \nonumber \\
= & 4 \cos ( { 1 \over 2 } (k_3 -k_4) )
\left [ \cos ( {1 \over 2} (2 k_1 + k_3 + k_4 ) ) +
\cos ( {1 \over 2} (2 k_2 + k_3 + k_4 ) ) \right ] \nonumber \\
\sim &  
8 - 2 k_1^2 -2 k_2^2 -2 k_1 k_3 - 2 k_2 k_3 -2 k_3^2 -2 k_1 k_4
-2 k_2 k_4 -2 k_4^2 +O(k^4)
\eea
To show the equivalence of the Regge action to the continuum Einstein
action one needs to replace the $e$ fields with metric
components (or alternatively, as done in \cite{rowi}, use
trace reversed metric components), with body principals expanded as
\bea
e_1 \; = \; -1 + \left [ 1+ \omega_1 h_{11} \right ]^{1/2} \nonumber \\
e_2 \; = \; -1 + \left [ 1+ \omega_2 h_{22} \right ]^{1/2} \nonumber \\
e_4 \; = \; -1 + \left [ 1+ \omega_4 h_{33} \right ]^{1/2} \nonumber \\
e_8 \; = \; -1 + \left [ 1+ \omega_8 h_{44} \right ]^{1/2} \;\; ,
\eea
face diagonals as
\bea
e_3 \; = \; -1 + \left [ 1 + 
{1 \over 2} \omega_1 \omega_2 (h_{11} + h_{22}) + h_{12} ) \right ]^{1/2}
\nonumber \\
e_5 \; = \; -1 + \left [ 1 + 
{1 \over 2} \omega_1 \omega_4 (h_{11} + h_{33}) + h_{13} ) \right ]^{1/2}
\nonumber \\
e_9 \; = \; -1 + \left [ 1 + 
{1 \over 2} \omega_1 \omega_8 (h_{11} + h_{44}) + h_{14} ) \right ]^{1/2}
\nonumber \\
e_6 \; = \; -1 + \left [ 1 + 
{1 \over 2} \omega_2 \omega_4 (h_{22} + h_{33}) + h_{23} ) \right ]^{1/2}
\nonumber \\
e_{10} \; = \; -1 + \left [ 1 + 
{1 \over 2} \omega_2 \omega_8 (h_{22} + h_{44}) + h_{24} ) \right ]^{1/2}
\nonumber \\
e_{12} \; = \; -1 + \left [ 1 + 
{1 \over 2} \omega_4 \omega_8 (h_{33} + h_{44}) + h_{34} ) \right ]^{1/2} \;\; ,
\eea
body diagonals as
\bea
e_7 \; = \; -1 + \left [ 1 + 
{1 \over 3} \omega_1 \omega_2 \omega_4 (h_{11} + h_{22} + h_{33}) +
{1 \over 3} ((1+\omega_4) h_{12} + (1+\omega_1) h_{23} + (1+\omega_2) h_{13})
 ) \right ]^{1/2} \nonumber \\
e_{11} \; = \; -1 + \left [ 1 + 
{1 \over 3} \omega_1 \omega_2 \omega_8 (h_{11} + h_{22} + h_{44}) +
{1 \over 3} ((1+\omega_8) h_{12} + (1+\omega_1) h_{24} + (1+\omega_2) h_{14})
 ) \right ]^{1/2} \nonumber \\
e_{13} \; = \; -1 + \left [ 1 + 
{1 \over 3} \omega_1 \omega_2 \omega_4 (h_{11} + h_{33} + h_{44}) +
{1 \over 3} ((1+\omega_8) h_{13} + (1+\omega_1) h_{34} + (1+\omega_4) h_{14})
 ) \right ]^{1/2} \nonumber \\
e_{14} \; = \; -1 + \left [ 1 + 
{1 \over 3} \omega_2 \omega_4 \omega_8 (h_{22} + h_{33} + h_{44}) +
{1 \over 3} ((1+\omega_8) h_{23} + (1+\omega_2) h_{34} + (1+\omega_4) h_{24})
 ) \right ]^{1/2} \nonumber \\
\eea
and finally hyperbody diagonals as
\bea
e_{15} \; = \; -1 + \left [ 1 + 
{1 \over 4} \omega_1 \omega_2 \omega_4 \omega_8 
( h_{11} + h_{22} + h_{33} + h_{44} ) +
{3 \over 4} ( h_{12} + h_{13} + h_{14} + h_{23} + h_{24} + h_{34} )
\right ]^{1/2} \;\; ,
\label{eq:etoh}
\eea
although the latter quantity is not needed, as it does not appear in the Regge
action to lowest order in the weak field expansion.
Each expression is then expanded out for weak $h$, giving for example
\bea
e_1 \; & = & \; { 1 \over 2 } \; \omega_1 h_{11} + O( h^2) \nonumber \\
e_3 \; & = & \; { 1 \over 2 } \; h_{12} + { 1 \over 4 } \; \omega_1 \omega_2
( h_{11} +  h_{22} ) + O( h^2)  \nonumber \\
e_7 \; & = & \; { 1 \over 6 } \; ( h_{12} + h_{13} + h_{23} ) +
{ 1 \over 6 } \; ( \omega_1 h_{23} + \omega_2 h_{13} + \omega_4 h_{12} ) +
{ 1 \over 6 } \; \omega_1 \omega_2 \omega_4 ( h_{11} +  h_{22} + h_{33} )
+ O( h^2)  \nonumber \\
\eea
and so on for the other edges by permuting indices.
Setting then $ \omega_1 = e^{i k_1}$ ... $ \omega_8 = e^{i k_4} $
(we switch here from the binary notation for the $\omega$'s to 
a normal notation for the $k$'s),
the resulting answer is finally expanded out in $k$ to give exactly
the weak field expansion of the continuum Einstein action
as given in Eq.~(\ref{eq:wfe}) and Eq.~(\ref{eq:wfe1}), and
completely parallels the procedure for recovering the
continuum limit of the lattice action as described in \cite{rowi,hw3d}.

To derive a lattice expression for the ground state functional
of linearized gravity one needs to compute the lattice
gravitational action on a given time slice, and subsequently impose the
appropriate discrete vacuum gauge conditions.
This will then give the action contribution appearing in the exponent of the 
ground state functional for linearized gravity as 
it appears in Eq.~(\ref{eq:kuchar}).

The first step involves therefore the imposition of the vacuum gauge 
conditions $ h_{ik,k}=0 $, $ h_{00}=h_{0i}=0 $ which gives
\bea
e_8  & = & 0 \nonumber \\
e_9  & = &  { 1 \over 2}  \; \omega_8 \; e_1 \nonumber \\
e_{10} & = & { 1 \over 2} \; \omega_8 \; e_2 \nonumber \\
e_{12} & = & { 1 \over 2} \; \omega_8 \; e_4 \nonumber \\
e_{11} & = & { 1 \over 3} \;  (1 + \omega_8 ) e_3 - 
\; {1 \over 6} \; (1 - \omega_8 ) ( \omega_2 e_1 + \omega_1 e_2 ) \nonumber \\
e_{13} & = & { 1 \over 3} \; (1 + \omega_8 ) e_5 - 
\; {1 \over 6} \; (1 - \omega_8 ) ( \omega_4 e_1 + \omega_1 e_4 ) \nonumber \\
e_{14} & = & { 1 \over 3} \; (1 + \omega_8 ) e_6 - 
 \; {1 \over 6} \; (1 - \omega_8 ) ( \omega_2 e_4 + \omega_4 e_2 )
\eea
and results in an action contribution of the form
\bea
& & 2 e_1^2 + 8 e_3^2 + ( \omega_1 \omega_4 + \omega_2 \omega_4 + \bar \omega_1 \bar \omega_4 + \bar \omega_2 + \bar \omega_4) e_1 e_2 
- 2 ( \omega_2 + \bar \omega_2 + 2) e_1 e_3 
\nonumber \\
& & - 2 ( \omega_1 \omega_2 + \bar \omega_1 \bar \omega_2 + \omega_4 + \bar \omega_4 ) e_3 e_4 
+  4 ( \omega_2 + \bar \omega_2 + \omega_4 + \bar \omega_4 ) e_3 e_5 + \cdots 
\label{eq:ttact}
\eea
where the dots indicate again additional terms obtainable by permutation of indices.

To verify that this is indeed the correct expression one can use the
expansion of the $e_i$'s in terms of the $h_{ij}$'s, as given in 
Eq.~(\ref{eq:etoh}), and then expand out the $\omega$'s in powers of $k$. 
One obtains
\bea
& & {1 \over 2} ( k_1^2 h_{23}^2 - k_1^2 h_{22} h_{33} - 2 k_1 k_2 h_{13} h_{23}
+ 2 k_1 k_2 h_{12} h_{33} + k_2^2 h_{13}^2 - k_2^2 h_{11} h_{33}  
+ 2 k_1 k_3 h_{13} h_{22} 
\nonumber \\
& & - 2 k_1 k_3 h_{12} h_{23} 
- 2 k_2 k_3 h_{12} h_{13} + 2 k_2 k_3 h_{11} h_{23} + k_3^2 h_{12}^2
- k_3^2 h_{11} h_{22} )
\eea
which can in turn be re-written as the sum of two parts, the first part being
the transverse-traceless contribution
\beq
{1 \over 4} {\bf k}^2
\Tr [ \; {}^3 h (P \; {}^3 h P - {1 \over 2} P \Tr (P \; {}^3 h )) ]
\; = \; {1 \over 4} {\bf k}^2 \bar h_{ij}^{TT} ({\bf k}) \; h^{TT}_{ij} ({\bf k})
\eeq
\beq
\bar h^{TT}_{ij} h^{TT}_{ij}  \; = \;
\Tr [ \; {}^3 h (P \; {}^3 h P - {1 \over 2} P \Tr (P \; {}^3 h )) ]
\eeq
with $P_{ij} = \delta_{ij} - k_i k_j / {\bf k}^2 $,
and the second part arising due to the trace component of the metric
\beq
-  { 1 \over 4 } {\bf k}^2
\Tr [ P \Tr (P \; {}^3 h) P \Tr (P \; {}^3 h ) ]
\; = \; {\bf k}^2 \bar h_{ij}^{T} ({\bf k}) \; h^{T}_{ij} ({\bf k})
\eeq
with $h^T= {1 \over 2} P \Tr (P \; {}^3 h) $. 
In the vacuum gauge $ h_{ik,k}=0 $, $ h_{ii}=0 $, $ h_{0i}=0 $ 
one needs to further solve for the metric components 
$h_{12}$, $h_{13}$, $h_{23}$ and $h_{33}$
in terms of the two remaining degrees of freedom, $h_{11}$ and $h_{22}$, 
\bea
h_{12} & = & - { 1 \over  2 k_1 k_2 } 
( h_{11} k_1^2 + h_{22} k_2^2 + h_{11} k_3^2 + h_{22} k_3^2 ) \nonumber \\
h_{13} & = & - { 1 \over  2 k_1 k_3 } 
( h_{11} k_1^2 - h_{22} k_2^2 - h_{11} k_3^2 - h_{22} k_3^2 ) \nonumber \\
h_{23} & = & - { 1 \over  2 k_2 k_3 } 
( - h_{11} k_1^2 + h_{22} k_2^2 - h_{11} k_3^2 - h_{22} k_3^2 ) \nonumber \\
h_{33} & = & - h_{11} - h_{22} 
\eea
and show that the second (trace) part vanishes.
For example, in terms of the $e$ variables the vacuum gauge condition
$ h_{ik,k}=0 $ then reads
\beq
\left [ 2 (1- {\bar \omega_1} )
+ \omega_2 (1- \omega_2) + \omega_4 (1-\omega_4)
\right ] e_1 + \omega_1 ( 1 - \omega_2) e_2 + \omega_1 ( 1 - \omega_4 ) e_4
- 2 ( 1 - \omega_2) e_3 - 2 (1-\omega_4 ) e_5 \; = \; 0
\label{eq:gauge}
\eeq
and permutations.
The above manipulations then show that the expression
given in Eq.~(\ref{eq:ttact}) is indeed the sought-after lattice analog for
the continuum expression $ {\bf k}^2 h_{ij}^{TT} \; \bar h^{TT}_{ij} $
appearing in the exponent of the ground state functional of lineraized
gravity.

We conclude this section by outlining an example of a potentially useful application
for the above results.
The explicit construction of the ground state wave functional
of linearized lattice gravity in terms of lattice transverse-traceless
modes makes it possible at least in principle to compare
the lattice and continuum results in the limit of small
curvatures, such as would be obtained for example from lattice simulations by imposing
flat boundary conditions at spatial infinity.
After imposing the boundary conditions by suitably restricting
the values for the edge lengths on the lattice
boundary such that the deficit angle is zero there,
one would then have to further enforce the lattice vacuum
gauge conditions of Eq.~(\ref{eq:gauge}) so as to finally make contact
with the semiclassical lattice functional of Eq.~(\ref{eq:ttact}).
But no gauge fixing is required for determining invariant averages obtained via
the partition function of Eq.~(\ref{eq:zlatt}), so
in practice the gauge conditions would have to be imposed 
configuration by configuration, by progressively
applying local gauge transformations \cite{gauge} so
as to gradually transform the edge lengths for each configuration
to the lattice analog of the vacuum gauge. 
It is expected that after such a transformation the edge distributions
on a fixed time slice should follow closely the distribution
of Eq.~(\ref{eq:ttact}), if indeed as expected the only surviving physical modes
are transverse traceless.

\vskip 30pt
\newsection{Ground State Wave Functional for Linearized Regge Calculus}
\hspace*{\parindent}

In the previous section the vacuum functional for linearized gravity
was derived by evaluating the discrete Euclidean action on a fixed
time slice, which was later supplemented by the appropriate gauge conditions.
In this section the ground state functional will instead be derived
by performing directly the discrete functional integration over
the interior metric perturbations for a lattice with boundary.

One of the main results of this paper is the analog of Hartle's
continuum calculation of the ground state wave functional
of linearized gravity, using Regge calculus.
We shall now briefly outline Hartle's
calculation since our calculation later is (intended to be) a discrete
version of his.

In linearized gravity, the Einstein action is expanded to quadratic
order in deviations of the metric from its flat-space value. On a
surface which becomes the flat surface $t = constant$ when the metric
perturbations are zero, we write the three-metric as 

\begin{equation}
{^3g_{ij} = \delta_{ij} + h_{ij}},
\end{equation}

\noindent and $h_{ij}$ can be decomposed into a transverse traceless part, a
longitudinal part and the trace. Since the physical degrees of freedom
are the two independent components of $h^{TT}_{ij}$, the transverse
traceless part, the wave function on the surface can be written as

\begin{equation}
{\Psi_0 = \Psi_0 [ h^{TT}_{ij}({\bf x}),t ] }.
\end{equation}

\noindent The Euclidean Einstein action is given by 

\begin{equation}
{l_P^2 I(g) \; = \; 
- \int_M d^4 x \sqrt{g} \; R - 2 \int_{{\partial}M} d^3 x \sqrt{^3g} \; K },
\end{equation}

\noindent and for linearised gravity, the Euclidean four-metric in the
functional integral is written as

\begin{equation}
{g_{\alpha\beta}(x) = \delta_{\alpha\beta} + h_{\alpha\beta}(x)}
\end{equation}

\noindent and the action is expanded to quadratic order in
$h_{\alpha\beta}$. The boundary $\partial M$ is taken to be a flat slice in
flat Euclidean space, and $M$ is the region of flat Euclidean space to
the past of this. The $h_{\alpha\beta}$ are required to vanish at
infinity so that $g_{\alpha\beta}$ is asymptotically flat.

The action is required to be gauge invariant, and gauge-fixing terms
in the four-volume and on the surface are included in the functional
integral. The metric perturbations are divided into conformal
equivalence classes \cite{hawking} by writing

\begin{equation}
{h_{\alpha\beta}(x) = \phi_{\alpha\beta}(x) + 2 \; \delta_{\alpha\beta} \;
\chi(x)}.
\end{equation}

\noindent
The integration contour for $\chi$ is rotated to purely imaginary
values to make the integral over $\chi$ converge. Then
the field $\phi_{\alpha\beta}$ is decomposed as

\begin{equation}
{\phi_{\alpha\beta} = \hat\phi_{\alpha\beta} + f_{\alpha\beta}},
\end{equation}

\noindent where $ \hat\phi_{\alpha\beta}$ is a solution of the linearised field
equations which satisfies the gauge and boundary conditions. The
unique solution is essentially that the spatial components are the
$h^{TT}_{ij}$ and the other components vanish. The integral over the
$f_{\alpha\beta}$ contributes only to the normalisation factor, as
does that over $\chi$. The result for the ground state wave function
of linearised gravity is

\begin{equation}
\Psi_0 [ h^{TT}_{ij} ,t ] =
{\cal N} \exp \left ( - {1 \over 4 l_P^2 } \int d^3 {\bf k} \; \omega_k \;
h^{TT}_{ij}({\bf k}) \; \bar h^{TT}_{ij}({\bf k}) \right ) ,
\end{equation}

\noindent where $h^{TT}_{ij}( {\bf k} )$ is the Fourier transform of
$h^{TT}_{ij}( {\bf x} )$, $ \omega_k=| {\bf k} |$, ${\cal N}$ is a normalisation factor
and $l_P =\sqrt{16\pi G}$ is the Planck length.
This is exactly the same formula as that obtained by
Kucha\u r using canonical methods.

Linearized Regge calculus can be implemented
as the theory of the small fluctuations of edge lengths away from
their flat-space values, in a tessellation of four-dimensional space
using rectangular hypercubes subdivided into four-simplices. The
methods of subdivision and the notation are described in detail in earlier work
on linearised Regge calculus \cite{rowi}, the difference in this
case being that we have a four-dimensional Euclidean space with a flat
boundary. The binary notation in \cite{rowi}, which we shall also
use here, comes from interpreting lattice vectors $(0,0,0,1),
(0,0,1,0), (0,1,0,0)$ and $(1,0,0,0)$ as binary numbers, giving the
$x_1$-, $x_2$-, $x_4$- and $x_8$-directions. For ease of notation here,
we shall take the boundary surface to be $x_8=0$, and the Euclidean
four-space to be $x_8\ge 0$ (to avoid lots of minus signs). Unlike the
continuum case, we shall take periodic boundary conditions in the 1-,
2- and 4-directions, while the space will be asymptotically flat in
the 8-direction. With unit lattice spacing, the flat-space edge
lengths will be $1, \sqrt{2}, \sqrt{3}$ and $2$, and the perturbed
edge lengths will be written as in Section 2, as

\begin{equation}
{l^j_i = L^j_i(1 + e^j_i)},
\end{equation}

\noindent where $L$ is the flat-space edge length, $e$ is a small perturbation,
and in each case, the upper index $j$ denotes the lattice point at
which the edge is based and the lower index $i$ denotes the direction
of that edge (all in binary notation). (Note a small change in
notation from \cite{rowi} where the small perturbations were called
$\delta$.) Thus, for instance, the $e$'s based at the origin and lying
in the boundary hypersurface will be $e^0_1, e^0_2, e^0_3, e^0_4,
e^0_5, e^0_6, e^0_7$.

In brief, our method is to write down the action for the
semi-infinite four-dimensional space and to perform a functional
integral over the internal perturbations, leaving an expression in
terms of the $e$'s on the boundary. The aim is then to identify the
quadratic expression in the surface $e$'s with the discrete version of
the integral of $\-h^{TT}_{ij}({\bf k}) \bar h^{TTij}({\bf k})$. The
calculation is long and tedious so we shall give relatively little
detail, but enough for the reader to reproduce it if required!

\subsection*{Interior Terms}

Consider an interior vertex, which for simplicity we shall label as if
it were the origin. The $e$'s for the edges based at this vertex will
be $e^0_i$, $i=1,2,...,15$, and we first write the total quadratic
contribution (the first non-vanishing order) to the action, which
involves any of these $e$'s. This will arise from the Regge action
$\Sigma A_h \delta_h$, for the hypercube based at
the origin and from neighbouring hypercubes. This is given explicitly
in Appendix B in \cite{rowi}.

The next step is to differentiate the action with respect to each of
the $e^0_i$ in turn to obtain their classical equations of
motion. Below we give an example of an equation of motion of each
type, for $e^0_1$, $e^0_3$ and $e^0_7$ respectively:

\[
0 = 6e^0_1 - 4(e^0_3 + e^0_5 + e^0_6) - 2(e^1_6 + e^1_{10} +
e^1_{12}) - 4(e^{-2}_3 + e^{-4}_5 + e^{-8}_9)
\]
\[
+ 3(e^{-2}_7 +
e^{-2}_{11} + e^{-4}_7 + e^{-4}_{13} + e^{-8}_{11} + e^{-8}_{13}) +
e^3_4 + e^3_8 + e^5_2 + e^5_8 + e^9_2 + e^9_4
\]
\begin{equation}
{+ e^{-6}_2 + e^{-6}_4 -
2e^{-6}_6 + e^{-10}_2 + e^{-10}_8 - 2e^{-10}_{10} + e^{-12}_4 +
e^{-12}_8 - 2e^{-12}_{12}}
\end{equation}

\[
0 = 8e^0_3 - 2(e^0_1 + e^0_2) - 3(e^0_7 + e^0_{11}) + 2 (- e^1_2 +
e^1_6 + e^1_{10}) + 2(- e^2_1 + e^2_5 + e^2_9)
\]
\begin{equation}
{- e^3_4 - e^4_8 -
e^{-4}_4 - e^{-8}_8 + 2(e^{-4}_5 + e^{-4}_6 + e^{-8}_9 + e^{-8}_{10})
- 3(e^{-4}_7 + e^{-8}_{11})}
\end{equation}

\begin{equation}
{0 = 6e^0_7 - 2e^0_3 - 2e^0_5 - 2e^0_6 + e^1_2 + e^1_4 - 2e^1_6 +
e^2_1 + e^2_4 - 2e^2_5 + e^4_1 + e^4_2 - 2e^4_3}.
\end{equation}

\noindent All other equations of motion may be obtained by cyclic permutations
of the indices.

We introduce new integration variables $f^j_i$ by

\begin{equation}
{e^j_i = \hat e^j_i + f^j_i},
\end{equation}

\noindent where the $\hat e^j_i$ satisfy the equations of motion above. By
subtracting each of the $\hat e^j_i$ times the corresponding classical
equation of motion from the contribution to the action based at the
origin,  the $\hat e^j_i$ are completely
eliminated, leaving only Gaussian integrals over the $f^j_i$, which
contribute only to the normalization. The same feature appears in the
continuum. (Note that as in \cite{rowi}, cross terms of the form
$e^0_i e^j_k$, where $j$ is a neighbouring lattice point of the origin,
are assigned half to each of the lattice points involved.) This
elimination of the interior terms would seem to hold independently of
whether we impose periodic boundary conditions or asymptotic
flatness.

\subsection*{Boundary Terms}

Having integrated over all contributions to the action from interior
vertices, we are now left with the contributions assigned to vertices
on the $x_8=0$ boundary. These will consist not only of terms involving $e$'s
based at vertices on the boundary hypersurface, but also of
contributions from vertices one layer in but assigned (as explained
above) partly to the boundary layer.

Suppose now that the origin is back on the boundary hypersurface. The
total contribution to the action involving $e^0_i$ has more than 200
terms so will not be reproduced here. Clearly there are no terms
involving $e^{-j}_i$ with $j=8,9,...,14$ since the boundary is at
$x_8=0$. All $e^j_i$ terms with $i$ or $j = 8,9,...,14$ are as for
interior vertices, and most of the other terms are half their interior
values.

Recall that the $e$'s in the boundary three-surface ($e^0_i,
i=1,2,...,7$) are fixed but $e^0_i$ with $i=8,9,...,14$ are to be
varied and have exactly the same classical equations of motion as
before. We again simplify the action by subtracting $e^0_i$ times its
equation of motion, for each of these, and then eliminate $e^0_{11},
e^0_{13}$ and $e^0_{14}$ using their equations of motion. At the same
time, we eliminate $e^0_7$ using a constraint identical to its
equation of motion. This looks somewhat suspect, but the motivation
and justification are as follows. In three-dimensional linearised
Regge calculus performed in a manner completely analogous to the
four-dimensional case in \cite{rowi}, the $e^0_7$ mode is not
dynamical and satisfies a constraint which turns out to be identical
to its equation of motion; this reduces the number of variables to
six, the correct number. We apply this result to our three-dimensional
boundary hypersurface.

As a first step in linking our position-space representation of the
action to the momentum representation in the Kucha\u r-Hartle formula
\cite{{kuchar},{hartle}}, we take the Fourier transform in the 1-,
2- and 4-directions, which are those in which there are periodic
boundary conditions. (In the 8-direction, we have contributions from
only one other layer, the first interior one). The details are
explained in \cite{rowi} but there the complex nature of the Fourier
transforms was not taken into account. Here our convention is that
$e^0_i e^a_j$ transforms to 
$\omega_a {\tilde e}^0_i {\bar{\tilde e}}^0_j$, where
$a=1,2,4$ and $\omega_a=e^{ik_a}$. For consistency with this, the
linear expressions in the $e$'s in the equations of motion transform
slightly differently from in \cite{rowi}, with $e^a_i$ transforming
to ${\bar\omega}_a {\tilde e}^0_i$. To simplify the notation, we immediately
drop the tildes from the Fourier transforms and the superscripts $0$.

We have not yet eliminated $e_8, e_9, e_{10}$ and $e_{12}$; this is
not straightforward in the way it was for $e_{11}, e_{13}$ and
$e_{14}$ as their equations of motion are simultaneous equations for
the four $e$'s. For example, the Fourier transforms of those for $e_8$
and $e_9$, with $\alpha_i$ defined to be $1-\omega_i$, are

\[
e_8(2(|\alpha_1|^2 + |\alpha_2|^2 + |\alpha_4|^2) - \alpha_1
\bar\alpha_2  - \alpha_2 \bar\alpha_1 -
\alpha_1 \bar\alpha_4 - \alpha_4 \bar\alpha_1 - \alpha_2 \bar\alpha_4
- \alpha_4 \bar\alpha_2)
\]
\[
+ 2e_9( -|\alpha_2|^2 - |\alpha_4|^2 +
\alpha_1 \bar\alpha_2 + \alpha_1 \bar\alpha_4)
+ 2e_{10}(- |\alpha_1|^2 - |\alpha_4|^2 + \alpha_2 \bar\alpha_1 +
\alpha_2 \bar\alpha_4)
\]
\[
+ 2e_{12}(- |\alpha_1|^2 - |\alpha_2|^2 + \alpha_4 \bar\alpha_1 +
\alpha_4 \bar\alpha_2) 
\]
\[= e_1(2\alpha_1 + 2\alpha_2 + 2\alpha_4 - \bar\alpha_2 -
\bar\alpha_4 - 2\alpha_1\alpha_2 - 2\alpha_1\alpha_4 +
\alpha_1\bar\alpha_2 + \alpha_1\bar\alpha_4)
\]
\[
+ e_2(2\alpha_1 + 2\alpha_2 + 2\alpha_4 - \bar\alpha_1 -
\bar\alpha_4 - 2\alpha_1\alpha_2 - 2\alpha_2\alpha_4 +
\alpha_2\bar\alpha_1 + \alpha_2\bar\alpha_4)
\]
\[
+ e_4(2\alpha_1 + 2\alpha_2 + 2\alpha_4 - \bar\alpha_1 - \bar\alpha_2
- 2\alpha_1\alpha_4 - 2\alpha_2\alpha_4 + \alpha_4\bar\alpha_1 +
\alpha_4\bar\alpha_2) 
\]
\[
+ 2e_3(2\alpha_1\alpha_2 - \alpha_1 - \alpha_2) +
2e_5(2\alpha_1\alpha_4 - \alpha_1 - \alpha_4) + 2e_6(2\alpha_2\alpha_4
- \alpha_2 - \alpha_4)
\]
\[
+ e^8_1(- 2\alpha_1 - \alpha_2 - \alpha_4 + 2\bar\alpha_2 +
2\bar\alpha_4) + e^8_2(- 2\alpha_2 - \alpha_1 - \alpha_4 +
2\bar\alpha_1 + 2\bar\alpha_4)
\]
\[
+ e^8_4(- 2\alpha_4 - \alpha_1 - \alpha_2 + 2\bar\alpha_1 +
2\bar\alpha_2) 
\]
\begin{equation}
+ 2e^8_3(\alpha_1 + \alpha_2) + 2e^8_5(\alpha_1 + \alpha_4) +
2e^8_6(\alpha_2 + \alpha_4) ,
\end{equation}

\[
e_8(- |\alpha_2|^2 - |\alpha_4|^2 + \alpha_2\bar\alpha_1 +
\alpha_4\bar\alpha_1) 
+ 2e_9(|\alpha_2|^2 + |\alpha_4|^2) - 2e_{10} \alpha_2\bar\alpha_1 -
2e_{12}\alpha_4\bar\alpha_1
\]
\[
= e_1(\bar\alpha_2 + \bar\alpha_4) + e_2(2\bar\alpha_1 - \alpha_2 -
\alpha_2\bar\alpha_1) 
+ e_4(2\bar\alpha_1 - \alpha_4 - \alpha_1\bar\alpha_1) + 2e_3 \alpha_2
+ 2e_5\alpha_4
\]
\begin{equation}
{+ e^8_1(\alpha_2 + \alpha_4) + e^8_2(\alpha_2 - 2\bar\alpha_1) +
e^8_4(\alpha_4 - 2\bar\alpha_1) - 2e^8_3\alpha_2 - 2e^8_5\alpha_4},
\end{equation}

\noindent and those for $e_{10}$ and $e_{12}$ by cyclic permutations
of the indices.

These equations are not all independent. Adding the left-hand sides
gives zero, while adding the right-hand sides gives

\[
0 = 2(e_1(|\alpha_2|^2 + |\alpha_4|^2 - \alpha_1 \alpha_2 - \alpha_1
\alpha_4) + e_2(|\alpha_1|^2 + |\alpha_4|^2 - \alpha_1 \alpha_2 -
\alpha_2 \alpha_4)
\]
\begin{equation}
{+ e_4(|\alpha_1|^2 + |\alpha_2|^2 - \alpha_1
\alpha_4 - \alpha_2 \alpha_4) + 2e_3 \alpha_1 \alpha_2 + 2e_5 \alpha_1
\alpha_4 + 2e_6 \alpha_2 \alpha_4)}.
\end{equation}

\noindent This expression is precisely the three-dimensional scalar
curvature at the origin \cite{wheeler} and is constrained to be
zero. (The constraint on a three-dimensional hypersurface also
includes $(tr K)^2$ and $tr K^2$ terms but these terms are of higher
order in the $e$'s.) We can also show that $\alpha_1$ times the $e_9$
equation plus $\alpha_2$ times the $e_{10}$ equation plus $\alpha_4$
times the $e_{12}$ equation gives zero on the left-hand side but on
the right-hand side, it gives half the difference between the
three-dimensional curvature scalars at the origin and at vertex $8$,
and so again is zero. ($^3R$ is also zero at the interior vertex $8$
as the constraint is propagated into the bulk.)
\smallbreak
Thus we have a situation where only two of the equations are
independent, and where the consistency is guaranteed by the
constraint. To solve the equations symmetrically, we take

\begin{equation}
{\lambda = e_8},
\end{equation}

\begin{equation}
{\mu = \alpha_1 e_9 + \alpha_2 e_{10} + \alpha_4 e_{12}}
\end{equation}

\noindent to be arbitrary parameters. We then obtain

\begin{equation}
{e_9 = {1 \over{2\Sigma}} (B + \lambda X_{24} + 2\bar\alpha_1 \mu)},
\end{equation}

\begin{equation}
{e_{10} = {1 \over{2\Sigma}} (C + \lambda X_{14} + 2\bar\alpha_2 \mu)},
\end{equation}

\begin{equation}
{e_{12} = {1 \over{2\Sigma}} (D + \lambda X_{12} + 2\bar\alpha_4 \mu)},
\end{equation}

\noindent where $\Sigma=|\alpha_1|^2+|\alpha_2|^2+|\alpha_4|^2$ and
$X_{ij}=|\alpha_i|^2+|\alpha_j|^2-\alpha_i\bar\alpha_k-\alpha_j\bar\alpha_k$
with $i,j,k=1,2,4$ and $k \ne i,j$. $B,C$ and $D$ are the expressions on
the right-hand sides of the $e_9, e_{10}$ and $e_{12}$ equations. We
now substitute for $\bar e_9, \bar e_{10}$ and $\bar e_{12}$ in the
boundary action. The total coefficients obtained for $\bar \lambda$
and $\bar \mu$ are both multiples of $^3R$ at the origin and so
vanish.
\smallbreak
What remains is an extremely long expression. We write the coefficient
of $e_i$ in this as $W_i+U_i$ (for \lq\lq wanted" and \lq\lq
unwanted"!). We will now give the expressions for $i=1,3$, all others
being obtainable by cyclic permutations of the indices.

\[
4 \Sigma W_1 = [\bar e_1(|\alpha_2|^2 + |\alpha_4|^2 -
\alpha_2\bar\alpha_4 - \alpha_4\bar\alpha_2)
\]
\[
+ \bar e_2 (- |\alpha_4|^2 + 2\alpha_1\alpha_4 +
2\bar\alpha_2\bar\alpha_4 - \alpha_1\bar\alpha_2 -
\alpha_1\bar\alpha_4 - \alpha_4\bar\alpha_2)               
\]
\[
+ \bar e_4 (- |\alpha_2|^2 + 2\alpha_1\alpha_2 +
2\bar\alpha_2\bar\alpha_4 - \alpha_1\bar\alpha_2 -
\alpha_1\bar\alpha_4 - \alpha_2\bar\alpha_4)
\]
\[
+ 2 \bar e_3 (\alpha_4\bar\alpha_2 - |\alpha_4|^2) + 2
\bar e_5 (\alpha_2\bar\alpha_4 - |\alpha_2|^2)
+ 2 \bar e_6 (\alpha_1\bar\alpha_2 + \alpha_1\bar\alpha_4
- 2\bar\alpha_2\bar\alpha_4)](2 + \Sigma) 
\]
\[
- 2 \bar e^8_1 (|\alpha_2|^2 + |\alpha_4|^2 - \alpha_2\bar\alpha_4 -
\alpha_4\bar\alpha_2) 
\]
\[
- 2 \bar e^8_2 (- |\alpha_4|^2 + 2\alpha_1\alpha_4 +
2\bar\alpha_2\bar\alpha_4 - \alpha_1\bar\alpha_2 - \alpha_1\bar\alpha_4
- \alpha_4\bar\alpha_2)
\]
\[
- 2 \bar e^8_4 (- |\alpha_2|^2 + 2\alpha_1\alpha_2 +
2\bar\alpha_2\bar\alpha_4 - \alpha_1\bar\alpha_2 -
\alpha_1\bar\alpha_4 - \alpha_2\bar\alpha_4)
\]
\begin{equation}
{- 4 \bar e^8_3 (\alpha_4\bar\alpha_2 - |\alpha_4|^2) - 4 \bar e^8_5
(\alpha_2\bar\alpha_4 - |\alpha_2|^2) - 4 \bar e^8_6
(\alpha_1\bar\alpha_2 + \alpha_1\bar\alpha_4 - 2 \bar\alpha_2\bar\alpha_4)},
\end{equation}

\[
2 \Sigma W_3 = [\bar e_1 (\alpha_2\bar\alpha_4 - |\alpha_4|^2) + \bar
e_2 (\alpha_1\bar\alpha_4 - |\alpha_4|^2)
\]
\[+ \bar e_4 (\alpha_1\bar\alpha_4 + \alpha_2\bar\alpha_4 -
2\alpha_1\alpha_2)
+ 2 \bar e_3 |\alpha_4|^2 - 2 \bar e_5 \alpha_2\bar\alpha_4 - 2 \bar
e_6 \alpha_1\bar\alpha_4](2 + \Sigma)
\]
\[- 2 \bar e^8_1 (\alpha_2\bar\alpha_4 - |\alpha_4|^2) - 2 \bar e^8_2
(\alpha_1\bar\alpha_4 - |\alpha_4|^2)
\]
\begin{equation}
{- 2 \bar e^8_4 (\alpha_1\bar\alpha_4 + \alpha_2\bar\alpha_4 -
2\alpha_1\alpha_2) - 4 \bar e^8_3 |\alpha_4|^2 + 4 \bar e^8_5
\alpha_2\bar\alpha_4 + 4 \bar e^8_6 \alpha_1\bar\alpha_4},
\end{equation}

\[
\Sigma U_1 = \bar e_2 (|\alpha_1|^2 - \bar\alpha_2|\alpha_1|^2 +
\alpha_1\alpha_4 + \alpha_4\bar\alpha_2 - \alpha_1\alpha_4\bar\alpha_2
+ \bar\alpha_2^2)
\]
\[
+ \bar e_4 (|\alpha_1|^2 - \bar\alpha_4|\alpha_1|^2 +
\alpha_1\alpha_2 + \alpha_2\bar\alpha_4 - \alpha_1\alpha_2\bar\alpha_4
+ \bar\alpha_4^2)
\]
\[
+ 2 \bar e_3 \bar\alpha_1\bar\alpha_2 + 2 \bar e_5
\bar\alpha_1\bar\alpha_4 + 2 \bar e_6 \bar\alpha_2\bar\alpha_4
\]
\[
- \bar e^8_1 (\alpha_1\alpha_2 + \alpha_1\alpha_4 -
\bar\alpha_1\bar\alpha_2 - \bar\alpha_1\bar\alpha_4)
\]
\[
+ \bar e^8_2 (- |\alpha_1|^2 + |\alpha_2|^2 - \alpha_1\alpha_2 +
\bar\alpha_1\bar\alpha_2 - \alpha_1\alpha_4 +
\bar\alpha_2\bar\alpha_4)
\]
\[
+ \bar e^8_4 (- |\alpha_1|^2 + |\alpha_4|^2 - \alpha_1\alpha_4
+\bar\alpha_1\bar\alpha_4 - \alpha_1\alpha_2 +
\bar\alpha_2\bar\alpha_4)
\]
\begin{equation}
{- 2 \bar e^8_3 \bar\alpha_1\bar\alpha_2 - 2  \bar e^8_5
\bar\alpha_1\bar\alpha_4 - 2  \bar e^8_6 \bar\alpha_2\bar\alpha_4},
\end{equation}

\begin{equation}
{\Sigma U_3 = 2(\bar e_1 \alpha_2\bar\alpha_1 + \bar e_2
\alpha_1\bar\alpha_2 - \bar e_4 \alpha_1\alpha_2(1 - \bar\alpha_4)
+ \bar e^8_1 \alpha_1\alpha_2 + \bar e^8_2 \alpha_1\alpha_2 +
\bar e^8_4 \alpha_1\alpha_2)}.
\end{equation}

\noindent We see that

\[
\Sigma U_1 = ( \; {}^3R(0) - \; {}^3R(8))/2 
\]
\[
- \bar e_1 (|\alpha_2|^2 + |\alpha_4|^2 - \bar\alpha_1\bar\alpha_2
- \bar\alpha_1\bar\alpha_2 - \bar\alpha_1\bar\alpha_4)
\]
\[
+ \bar e_2 (\bar\alpha_2^2 - |\alpha_4|^2 - \alpha_1\bar\alpha_2
+\alpha_1\alpha_4 + \alpha_4\bar\alpha_2 + \bar\alpha_2\bar\alpha_4 -
\alpha_1\alpha_4\bar\alpha_2)
\]
\[
+ \bar e_4 (\bar\alpha_4^2 - |\alpha_2|^2 -\alpha_1\bar\alpha_4 +
\alpha_1\alpha_2 + \alpha_2\bar\alpha_4 - \bar\alpha_2\bar\alpha_4 -
\alpha_1\alpha_2\bar\alpha_4)
\]
\begin{equation}
{+ (\bar e^8_1 + \bar e^8_2 + \bar e^8_4)(|\alpha_2|^2 + |\alpha_4|^2
- \alpha_1\alpha_2 - \alpha_1\alpha_4)}. 
\end{equation}

\noindent Repeatedly using $^3R(0)= \; ^3 \! R(8)=0$, we then have the
following expression for the boundary action:

\[
\Sigma ( e_1U_1 + e_2U_2 + e_3U_3 + e_4U_4 + e_5U_5 + e_6U_6) =
\]
\begin{equation}
{(\bar e^8_1 + \bar e^8_2 + \bar e^8_4 - (\bar e_1 + \bar e_2 +
\bar e_4)){^3R(0)}/2 + U},   
\end{equation}

\noindent where the remainder $U$ is given by

\[
U = [(e_1 [\bar e_1 (\bar\alpha_1\bar\alpha_2 - \alpha_1\alpha_2
+\bar\alpha_1\bar\alpha_4 - \alpha_1\alpha_4)
\]
\[
+ \bar e_2(|\alpha_2|^2 + \bar\alpha_2^2 - \alpha_1\alpha_2 - \alpha_1
\bar\alpha_2 + \alpha_4\bar\alpha_2 +\bar\alpha_2\bar\alpha_4 -
\alpha_1\alpha_4\bar\alpha_2)
\]
\[+ \bar e_4 (|\alpha_4|^2 + \bar\alpha_4^2 - \alpha_1\alpha_4 -
\alpha_1\bar\alpha_4 +\alpha_2\bar\alpha_4 + \bar\alpha_2\bar\alpha_4
- \alpha_1\alpha_2\bar\alpha_4)]
\]
\[
+ e_2 [\bar e_1 (|\alpha_1|^2 + \bar\alpha_1^2 - \alpha_2\bar\alpha_1
- \alpha_1\alpha_2 + \alpha_4\bar\alpha_1 + \bar\alpha_1\bar\alpha_4 -
\alpha_2\alpha_4\bar\alpha_1)
\]
\[
+ \bar e_2 (\bar\alpha_1\bar\alpha_2 - \alpha_1\alpha_2 +
\bar\alpha_2\bar\alpha_4 - \alpha_2\alpha_4)
\]
\[
+ \bar e_4 (|\alpha_4|^2 + \bar\alpha_4^2 - \alpha_2\bar\alpha_4 -
\alpha_2\alpha_4 + \alpha_1\bar\alpha_4 + \bar\alpha_1\bar\alpha_4 -
\alpha_1\alpha_2\bar\alpha_4]
\]
\[
+ e_4 [\bar e_1 (|\alpha_1|^2 + \bar\alpha_1^2 - \alpha_4\bar\alpha_1
- \alpha_1\alpha_4 + \alpha_2\bar\alpha_1 + \bar\alpha_1\bar\alpha_2 -
\alpha_2\alpha_4\bar\alpha_1)
\]
\[
+ \bar e_2 (|\alpha_2|^2 + \bar\alpha_2^2 - \alpha_4\bar\alpha_2 -
\alpha_2\alpha_4 + \alpha_1\bar\alpha_2 + \bar\alpha_1\bar\alpha_2
-\alpha_1\alpha_4\bar\alpha_2)
\]
\[
+ \bar e_4 (\bar\alpha_1\bar\alpha_4 - \alpha_1\alpha_4 +
\bar\alpha_2\bar\alpha_4 - \alpha_2\alpha_4)]
\]
\[
+ 2 e_3 [ \bar e_1 (\alpha_2\bar\alpha_1 + \alpha_1\alpha_2) + \bar
e_2 (\alpha_1\bar\alpha_2 + \alpha_1\alpha_2) + \bar e_4
\alpha_1\alpha_2\bar\alpha_4)] 
\]
\[
+ 2 e_5 [ \bar e_1 (\alpha_4\bar\alpha_1 + \alpha_1\alpha_4) + \bar
e_2 \alpha_1\alpha_4\bar\alpha_2 + \bar e_4 (\alpha_1\bar\alpha_4 +
\alpha_1\alpha_4)] 
\]
\begin{equation}
{+ 2 e_6 [ \bar e_1 \alpha_2\alpha_4\bar\alpha_1 + \bar e_2
(\alpha_4\bar\alpha_2 + \alpha_2\alpha_4) + \bar e_4
(\alpha_2\bar\alpha_4 + \alpha_2\alpha_4)]}
\end{equation}

\smallbreak
The next step is to expand in powers of $k$, using $\omega_i=e^{ik_i},
\; \alpha_i =1-\omega_i$ (in contrast to the previous
section, here we keep the binary notation for the $k_i$'s).
For the remainder $U$ above, we obtain

\begin{equation}
{U = 2 i (k_1 \bar e_1 + k_2 \bar e_2 + k_4 \bar e_4)\\{^3R(0)} +
O(k^4)}.
\end{equation}

\noindent For $W_1$ and $W_3$, we have

\[
2(k_1^2 + k_2^2 + k_4^2)W_1 = (\bar e_1 - \bar e^8_1)(k_2^2 + k_4^2
- 2k_2k_4) 
\]
\[
- (\bar e_2 - \bar e^8_2)(k_4^2 + 3k_1k_4 + 3k_2k_4 + k_1k_2) - (\bar
e_4 - \bar e^8_4)(k_2^2 + 3k_1k_2 + 3k_2k_4 + k_1k_4)
\]
\[
+ 2(\bar e_3 - \bar e^8_3)(k_2k_4 - k_4^2) + 2(\bar e_5 - \bar
e^8_5)(k_2k_4 - k_2^2)
\]
\begin{equation}
{+ 2(\bar e_6 - \bar e^8_6)(k_1k_2 + k_1k_4 + 2k_2k_4) + O(k^3)},
\end{equation}

\[
(k_1^2 + k_2^2 + k_4^2)W_3 = (\bar e_1 - \bar e^8_1)(k_2k_4 - k_4^2) +
(\bar e_2 - \bar e^8_2)(k_1k_4 - k_4^2)
\]
\[
+ (\bar e_4 - \bar e^8_4)(k_1k_4 + k_2k_4 + 2k_1k_2) + 2(\bar e_3 -
\bar e^8_3)k_4^2
\]
\begin{equation}
{- 2(\bar e_5 - \bar e^8_5)k_2k_4 - 2(\bar e_6 - \bar e^8_6)k_1k_4 + O(k^3)}.
\end{equation}

\noindent Thus our final expression for the boundary action is

\[
{1 \over {2(k_1^2 + k_2^2 + k_4^2)}}[e_1[(\bar e_1 - \bar
e^8_1)(k_2^2 + k_4^2 - 2k_2k_4)
\]
\[- (\bar e_2 - \bar e^8_2)(k_4^2 + 3k_1k_4 + 3k_2k_4 + k_1k_2) -
(\bar e_4 - \bar e^8_4)(k_2^2 + 3k_1k_2 + 3k_2k_4 + k_1k_4)
\]
\[
+ 2(\bar e_3 - \bar e^8_3)(k_2k_4 - k_4^2) + 2(\bar e_5 - \bar
e^8_5)(k_2k_4 - k_2^2)
\]
\[
+ 2(\bar e_6 - \bar e^8_6)(k_1k_2 + k_1k_4 + 2k_2k_4) + O(k^3)]
\]
\[
+ e_2 [...] + e_4 [...]
\]
\[
+ 2 e_3[(\bar e_1 - \bar e^8_1)(k_2k_4 - k_4^2) + (\bar e_2 - \bar
e^8_2)(k_1k_4 - k_4^2)
\]
\[
+ (\bar e_4 - \bar e^8_4)(k_1k_4 + k_2k_4 + 2k_1k_2) + 2(\bar e_3 -
\bar e^8_3)k_4^2 
\]
\[
- 2(\bar e_5 - \bar e^8_5)k_2k_4 - 2(\bar e_6 - \bar e^8_6)k_1k_4 +
O(k^3)]
\]
\begin{equation}
{+ 2 e_5[...] + 2 e_6[...]]}
\end{equation}

\noindent The coefficients of $e_2$ and $e_4$ can be obtained from
those of $e_1$, and those of $e_5$ and $e_6$ from those of $e_3$, by
cyclic permutation of indices.
\smallbreak
This is to be compared with the expression for
$h^{TT}_{ij} \bar h^{TTij}$ stated earlier, Eq.~(\ref{eq:ttact})
supplemented by the gauge conditions of Eq.~(\ref{eq:gauge}).
Note that that expression
does not distinguish between $\bar e_i e_j$ and $\bar e_j e_i$, and
once this is taken into account, the expressions are identical
(apart from an overall minus sign which arises from the
fact that the Regge calculus expressions are calculated
from the changes in dihedral angles rather than the deficit angles).
Thus we can write our expression for the action as

\begin{equation}
{I = \int d^3 {\bf k} \; h^{TT}_{ij} (\bar h^{TTij}(0) - \bar
h^{TTij}(8))}
\end{equation}

\noindent or

\begin{equation}
{{I = \int d^3 {\bf k} \; h^{TT}_{ij} (-{\partial \over {\partial z}} \bar
h^{TTij})} \; 
{= \int d^3 {\bf k} \; \omega_k \; h^{TT}_{ij} \bar h^{TTij}}}
\end{equation}

\noindent where $\omega_k=k_8=\sqrt{k_1^2+k_2^2+k_4^2}$. This is using
the behaviour of $h(k)$ in a space with Euclidean signature, where, if
the expansion is periodic in the 1-, 2-, and 4-directions, we
shall have

\begin{equation}
{h(k) = C \exp \left [ i(k_1x_1 + k_2x_2 + k_4x_4) - k_8x_8 \right ] },
\end{equation}

\noindent with $k_8^2=k_1^2+k_2^2+k_4^2$ to satisfy the wave equation
$\Box h=0$.
\smallbreak
The ground state wave function obtained is thus identical to the
continuum result of Kucha\u r and Hartle.

\vskip 30pt
\newsection{The Evidence for Spin Two}
\hspace*{\parindent}

In the weak field limit the lattice theory described by the 
partition function Eq.~(\ref{eq:zlatt}) is known to be equivalent to
the continuum theory of a massless spin two particle, as embodied in Einstein's General
Relativity with a cosmological constant term.
One would hope that the local gauge invariance (continuous lattice diffeomorphism
invariance) of the discrete gravitational action under metric deformations
- taken sufficiently small so as not to violate the triangle
inequalities \cite{gauge} - would be powerful enough to ensure that the lattice
theory still describes a regularized model for quantum gravity, even
away from smooth manifolds. 
In this section we shall examine the evidence in support of the
argument that the lattice theory, treated non-perturbatively
and in the vicinity of the critical
point at $G_c$ where the lattice continuum limit is formally defined,
still describes a massless spin two particle.
A comparison will be made between the lattice results and those
obtained recently in the continuum using a variety of perturbative ($2+\epsilon$
expansion) and non-perturbative (renormalization group combined
with a derivative expansion) methods. 
A second line of approach will be to compare the lattice results
for the critical exponents of gravitation with what is known either exactly
or approximately for other spin values (0,1)
in four dimensions, and look for a discernible trend.

The cornerstone for these kind of arguments is the basic
idea of universality.
It is known that the long distance behavior of quantum field theories is
to a great extent determined by the scaling behavior of the coupling constant
under a change in momentum scale \cite{wilson}.
It is also well known that asymptotically
free theories such as QCD lead to vanishing gauge
couplings at short distances, while the opposite is
true for QED and self-interacting scalar field theories in four dimensions.
More generally, the fixed points of the renormalization
group need not be at zero coupling, but can be located at
some finite coupling, leading to non-trivial fixed points
not necessarily accessible by perturbation theory,
or possibly even more complex fixed lines and limit cycles \cite{wilson,gross}.

The existence of non-trivial ultraviolet fixed points,
at which the theory becomes scale invariant,
corresponds in statistical mechanics language
to the existence of one or more critical points.
There the partition function exhibits non-analyticities
and singularities caused by infrared divergences
associated with a divergent correlation length.
It has been shown \cite{lesh,critical} that lattice gravity
exhibits precisely such a transition
where, for example, the curvature fluctuation
\beq
\chi_{\cal R}  (k) \; \sim \;
{ < ( \int \sqrt{g} \, R )^2 > - < \int \sqrt{g} \, R >^2
\over < \int \sqrt{g} > } \;\; .
\eeq
diverges at some $k_c$.
Such a divergence signals a singularity in
the partition function itself, since averages such as the average curvature
${\cal R}$ and the curvature fluctuation
$\chi_{\cal R}$ are related to derivatives of $Z_L$
(of Eq.~(\ref{eq:zlatt})),
with respect to the gravitational coupling $k=1/(16 \pi G)$.

Simple scaling arguments allow one to determine the scaling
behavior of correlation functions from the critical exponents which
characterize the singular behavior of local averages in the 
vicinity of the critical point.
The appearance of a singularity in the free energy $F(k)$ is
caused by the divergence of the correlation length $\xi $, which close
to the critical point at $k_c$ is assumed to behave as
\beq
\xi \; \equiv \; 1 / m \;
\mathrel{\mathop\sim_{ k \rightarrow k_c}} \; A_\xi \;
( k_c - k ) ^{ -\nu } \;\; 
\label{eq:m_latt}
\eeq
and defines the exponent $\nu$.
Since for the singular part of the free energy one expects $F_{sing}(k) \sim \xi^{-d}$
simply on dimensional grounds, one then obtains by differentiation
with respect to $k$ for the curvature fluctuation
\beq
\chi_{\cal R} (k) \; \mathrel{\mathop\sim_{ k \rightarrow k_c}} \;
A_{\chi_{\cal R}} \; ( k_c - k ) ^{ -(2- d \nu ) } \;\;\; .
\label{eq:chising}
\eeq
The last expression allows, at least in principle, a direct determination
of the critical exponent $\nu$.
Large scale direct numerical studies of the lattice theory \cite{critical}
give the value $ \nu = 0.335(9) $ for $ G_c=0.626(11) $,
which suggests $\nu = 1/3$ for pure gravitation.

Apart from a detailed comparison between critical exponents
(which will be done later in this paper),
a number of direct and indirect arguments can be given in support
of the fact that the non-perturbative lattice theory still
describes a massless spin two particle in the vicinity
of the critical point at $k_c$.
Firstly the gravitational lattice action only propagates
spin two (transverse traceless) degrees of freedom, as shown explicitly in the
weak field expansion of the previous sections (the lattice
gravitational functional measure is completely local, and does
not contain any propagation terms, to any order of the weak field
expansion).
This result is further supported by rigorous work describing
the convergence of the lattice action towards the continuum
one for smooth enough manifolds \cite{cms,lee}.
Secondly, the static interaction of two heavy particles of mass $m$
described by two world lines kept at a fixed distance $d$
has been shown to scale consistently as the mass squared, as expected for gravitational
type interactions \cite{lines}.
In the following we will explore this delicate issue further, by pursuing
the connection with available non-perturbative results in the continuum.

\newpage

\vskip 30pt
\subsection*{Ultraviolet Fixed Point}
\hspace*{\parindent}

One can contrast and compare the lattice results with what one
obtains for quantum gravity in the continuum.
Since gravity is not perturbatively renormalizable in
four dimensions, one has to go to a lower dimension 
$(d=2)$ where the perturbative expansion becomes meaningful,
and expand about that dimension. 
Similar expansion have been shown to be quantitatively very successful
in scalar field theories \cite{zj,zjg}, but the series are shorter
and a significantly larger
extrapolation is required in the gravitational case
(for a general review of the diagrammatic field theory methods as applied to statistical
mechanics models see \cite{pbook,izbook,zjbook}).

In the $2 + \epsilon$ perturbative expansion for gravity
\cite{epsilon} (earlier references can be found in \cite{epsilon1,epsilon2,epsilon3})
one analytically continues in the spacetime dimension by using
dimensional regularization, and applies perturbation theory about
$d=2$, where the theory is formally power counting renormalizable 
and Newton's constant is dimensionless.
An expansion in the number of dimensions of course goes back
to Wilson's original work \cite{wilson}, and since then
similar methods have been shown
to be quite successful in determining among others the critical properties
of the $O(n)$-symmetric non-linear sigma model above two
dimensions \cite{brezin}.
This model is not
perturbatively renormalizable either, yet describes a completely well-defined
and physically relevant statistical spin system, namely the
universality class of the 3-d Heisenberg ferromagnet.
The same dimensional expansion methods have been extended with some success to
fermionic models as well \cite{gn}.

In the gravitational case the dimensionful bare coupling is written as
$G_0 = \Lambda^{2-d} G $, where $G$ is dimensionless and $\Lambda$
is an ultraviolet cutoff (not to be confused here with the
scaled cosmological constant), corresponding on the lattice to a momentum
cutoff of the order of the inverse average
lattice spacing, $\Lambda \sim 1/ l_0 $.
The method has of course its share of problems, as the Einstein action
is a topological invariant in two dimensions, which leads
to kinematic singularities in the propagator.
In addition, to recover
the physical case d=4 requires a rather bold extrapolation from two dimensions.
The series themselves are rather short and strong assumptions need
to be made about the nature of possible singularities in the complex coupling
constant plane (in particular the absence of singularites close
to d=3).
Still, one can view this approach as providing some sort
of gauge-invariant resummation
of a specific set of subdiagrams which may or may not be 
ultimately relevant in d=4.

A double expansion in $G$ and $\epsilon= d-2$
then leads above two dimensions to a non-vanishing
beta function
\beq
\beta (G) \, \equiv \, { \partial G \over \partial \log \Lambda } \, = \,
(d-2) \, G \, - \, \beta_0 \, G^2 \, - \, \beta_1 \, G^3 \, 
+ \cdots \;\; ,
\label{eq:beta} 
\eeq
and consequently a nontrivial ultraviolet fixed point in $G$, since
$\beta_0 > 0 $ for pure gravity.
Integrating Eq.~(\ref{eq:beta}) close to the fixed point,
one obtains for $G > G_c $ a non-perturbative, dynamically generated
mass scale
\beq
m \, = \, \Lambda \, \exp \left ( { - \int^G \, {d G' \over \beta (G') } }
\right )
\, \mathrel{\mathop\sim_{G \rightarrow G_c }} \,
\Lambda \, | \, G - G_c |^{ - 1 / \beta ' (G_c) } \;\;\; .
\label{eq:m_cont}
\eeq
It should be noted at this point that Eq.~(\ref{eq:m_cont}) is essentially
the same as Eq.~(\ref{eq:m_latt}), with slightly
different notations.
It also brings out the central importance
of the exponent $\nu$, and how it relates to the scale
dependence of the coupling $G$. 
The derivative of the beta function at the fixed point defines
the critical exponent $\nu$ (which to lowest order is in fact
independent of $\beta_0$),
\beq
\beta ' (G_c) \, = \, - 1/ \nu  \;\; .
\eeq
In the previous expression $m$ is an arbitrary integration constant,
with dimensions of a mass, and has to be associated with some physical scale
to be determined (as in QCD) by physical considerations (we will argue
that it is the analog of $\Lambda_{\overline{MS}}$ for gravitation).
It would appear natural here to identify it with the inverse of
a gravitational correlation length ($\xi=m^{-1}$),
perhaps a length scale  associated with some average long
distance curvature (more on this later).
The above renormalization group
result also illustrates in a direct way how the lattice
continuum limit should be taken.
It corresponds to taking the ultraviolet cutoff $\Lambda \rightarrow \infty$,
and therefore $G \rightarrow G_c$, with $m$ held constant.
For a fixed lattice cutoff, the continuum limit is approached
by tuning $G$ to $G_c$.

The value of the universal critical exponent $\nu$ has important physical
consequences, as it directly determines the
running of the effective coupling $G(\mu)$, where $\mu$ is an
arbitrary momentum scale. 
The renormalization group tells us
that in general the effective coupling will
grow or decrease with length scale $ r = 1/ \mu$, depending on whether
$G > G_c$ or $G < G_c$, respectively.
The physical mass parameter $m$ is itself by definition scale independent,
and therefore obeys a Callan-Symanzik renormalization group equation,
which in the immediate vicinity of the fixed point takes on the simple form
\beq
\mu \; { \partial \over \partial \mu } \; m ( \mu) \; = \;
\mu \; { \partial \over \partial \mu } \;
\left \{ \; A_m \, \mu \, | \, G (\mu) - G_c |^{ \nu } \; \right \}
\, = \, 0 
\label{eq:callan} 
\eeq
with $A_m$ a numerical constant.
As a consequence,
for $G > G_c$, corresponding to the smooth phase, one expects for the
running, effective gravitational coupling \cite{critical,det}
\beq
G(r) \; = \; G(0) \left [ \; 1 \, + \, c \, ( r / \xi )^{1 / \nu} \, 
+ \, O (( r / \xi )^{2 / \nu} ) \; \right ] \;\; ,
\label{eq:grun1} 
\eeq
with $c$ a calculable numerical constant of order one.
\footnote{At very short distances $r\sim l_P$ one finds
finite perturbative corrections to the potential as well, which can be computed
analytically using weak coupling diagrammatic techniques \cite{donoghue}. }
The physical renormalization group invariant mass $m = \xi^{-1}$ determines the magnitude of scaling
corrections, and separates the short distance, ultraviolet regime 
from the large distance, infrared region.
As already mentioned in the introduction,
there are in fact indications that in the Euclidean lattice theory only the smooth
phase with $G>G_c$ exists (since spacetime becomes branched-polymer like and
therefore degenerate
for $G<G_c$), which would then imply that the 
gravitational coupling can only {\it increase} with distance
(this point will be discussed further in Section 5).

Recently the continuum $2+\epsilon$ expansion for gravitation
has been pushed to two loops, giving close to two dimensions
\cite{epsilon}
\beq
\beta (G) \, = \, (d-2) \, G \, - \, { 2 \over 3 } (25- n_f) \, G^2 \,
- \, { 20 \over 3 } (25- n_f) \, G^3 \, + \cdots \;\; ,
\label{eq:betaeps}
\eeq
for $n_f$ massless real scalar fields minimally coupled to gravity.
After solving the equation $\beta (G_c) \, = \, 0 $ to establish the location
of the fixed point, one obtains for pure gravity ($n_f=0$) 
\beq
G_c= (3/50) (d-2) - (9/250) (d-2)^2 + \cdots 
\eeq
and therefore close to two dimensions
\beq
\nu^{-1} \, = \, - \beta ' (G_c) \, =  \;\;
(d-2) \, + \, {3 \over 5} (d-2)^2 \, + O (d-2)^3 \;\; ,
\label{eq:nueps}
\eeq
which gives to lowest order $\nu^{-1} = 2$ independently of $d$, 
and $\nu^{-1} = 4.4$ at the next order in $d=4$.
The uncertainty in these results can perhaps best be judged
by comparing to similar calculations in the scalar case, for which
much longer series exist, and for which rather sophisticated
resummation methods based on Pade-Borel transforms, conformal mappings,
and incorporating
asymptotic large order estimates, are available \cite{zj,zjg}
(the methods of statistical field theory are discussed in detail in \cite{pbook,izbook}).
Unfortunately in general the convergence properties of the $2+\epsilon$
expansion for the non-linear sigma model are not encouraging,
even when comparing to well-established results
in $d=3$ ($\epsilon=1$) \cite{brezin}.

The $2+\epsilon$ expansion is not the only method that
has been applied in the continuum to extract quantitative 
informations about non-perturbative properties of gravitation.
In this context we should mention another set of related results for
the critical exponents of quantum gravitation.
Recently in a separate, approximate renormalization group calculation
based on the Einstein-Hilbert action truncation \cite{litim}
one finds in the limit of vanishing bare cosmological constant 
$ \nu^{-1} \, = \, 2 d (d-2) / (d+2) = 2.667$ in $d=4$, 
and $\nu^{-1} \approx 1.667 $ in a more elaborate truncation.
In this paper the sensitivity of the results to the choice
of gauge fixing term and to the specific shape of the momentum cutoff
is investigated as well.
These more recent results extend earlier calculations for
the exponent $\nu$ done by similar operator truncation methods, and
described in detail in references \cite{reuters, reuterl}.
A quantitative comparison of these various continuum results with the lattice
answer for $\nu^{-1}$ will be postponed until later in this paper.

\vskip 30pt
\subsection*{Geometric Argument for ${\bf \nu=1/3}$}
\hspace*{\parindent}

A simple geometric argument can be given in support of the exact value
$\nu=1/3$ for pure quantum gravitation.
The vacuum polarization induced scale dependence of the gravitational
coupling $G(r)$ as given in Eq.~(\ref{eq:grun1}) implies
the following quantum corrected static gravitational potential,
for a point source of mass $M$ located at the origin,
\beq
V(r) \; = \; - \; G(r) \; { m M \over r } \; = \; - \; G(0) \; { m M \over r }
\left [ \; 1 + c \; ( r  / \xi )^{1 / \nu} + {\cal O} ( ( r / \xi)^{2 / \nu} ) \;
\right ]
\label{eq:vrun}
\eeq
and for intermediate distances $ l_p \ll r \ll \xi $. 
As a result, the vacuum polarization effects due to virtual graviton loops
cause an effective anti-screening of the primary gravitational source $M$.
Thus the effect of the running gravitational coupling $G(r)$ is to give rise to a new
non-perturbative quantum contribution to the potential, proportional to $r^{1/ \nu -1}$.
Remarkably for $\nu=1/3$ the additional contribution, now proportional to 
$r^2$, can be interpreted as being due to what ultimately appears as a uniform
mass distribution surrounding the original source.
Its origin lies with a non-perturbative graviton vacuum polarization
contribution, localized around the point source, and of strength
\beq
\rho_0 \; = \; { 3 c M  \over 4 \pi \xi^3 } \;\;\; .
\eeq
Of course such a simple geometric interpretation fails unless
the critical exponent $\nu$ for gravitation is exactly one third.
In fact in any dimensions
$d \ge 4$ one would expect based on the geometric argument
that $- \beta'(G_c) = \nu^{-1} = d-1 $, if the leading
correction to the gravitational potential is due to a uniformly
distributed, anti-screening cloud of virtual gravitons. 
These arguments rely of course on the lowest order result 
$V(r) \sim \int d^{d-1} p \; e^{i p \cdot x} / p^2
\sim r^{3-d}$ for single graviton exchange in $d>3$ dimensions. 

Equivalently, the running of $G$ can be characterized as being
due to a tiny non-vanishing (and positive)
non-perturbative gravitational vacuum contribution to the cosmological
constant, with
\beq
\lambda_\xi \; = \; { 3 c M \over \xi^3 }
\eeq
and therefore an associated effective curvature of magnitude
$ {\cal R} \sim G \lambda_\xi \sim G M / \xi^3 $.
It is amusing that for a very large mass distribution,
the above expression for the curvature can only be
reconciled with the naive dimensional estimate ${\cal R} \sim 1 / \xi^2 $,
provided for the gravitational coupling $G$ itself one has $G \sim \xi / M $
\cite{lense}.

\vskip 30pt
\subsection*{Random Gravitational Paths}
\hspace*{\parindent}

Within the Feynman path integral formulation of quantum
field theory, a well known relationship exists between
the properties of random paths and those
of field correlations (see for example \cite{polybook}).
In this section the analogy will be exploited in trying to
gain more insight on the specific values for the gravitational
critical exponents.

In the simpler case of self-interacting scalar field theories a
rigorous argument can be given \cite{aizenman}
based on an exact geometric characterization of
criticality in the $\lambda \phi^4$ theory and the Ising model,
in and above four dimensions.
The key element of the argument lies in the recognition
of the fact that random walks representing the propagation
of free particles in Euclidean space-time have fractal \cite{mandel}
dimensions $d_H= \nu^{-1} = 2 $, 
with vanishing probability of self-intersection above $d=4$.
As a consequence these models are governed by mean field
theory above $d=4$, with mild logarithmic corrections to free
field behavior at $d=4$.
They provide rigorous support for the original claim that
in the infinite cutoff limit all scalar field theories are
trivial in four dimensions \cite{wilson}.

Let us first illustrate these results for the simplest case of
a free scalar field in $d$ dimensions with action
\beq
S \; = \; \half \int_{x,y} \phi (x) M(x,y) \phi (y) \;\;\; .
\eeq
On a lattice one has for the matrix $M_{ij} = D_{ij}-S_{ij}$, where
$S$ is the (nearest-neighbor) hopping part, and the rest is the diagonal
part $D_{ij}=(2d+m_0^2) \; \delta_{ij}$, with $d$ the dimensions and $m_0$
the bare mass.
The propagator connecting point $1$ to point $2$ is then given in terms
of the kernel $S$ by
\beq
G_{12} \; = \; {1 \over m_0^2 + 2 d } 
\sum_{n=0}^{\infty} { S \over m_0^2 + 2 d }
\eeq
or, equivalently, in terms of a sum over paths
\beq
G_{12} \; = \; \sum_{{\rm paths} \, 1 \rightarrow 2 } 
e^{- m \, l_{12} ({\rm path}) }
\; \sim \; r_{12}^{-(d-1)/2} \; e^{- r_{12} / \xi} 
\label{eq:corr_rw}
\eeq
where $l_{12} ({\rm path})$ is the length of the random path connecting
points $1$ and $2$. 
In the second part of the expression we
have indicated the asymptotic behavior of the free propagator
for large distances,
which brings in the correlation length $\xi = m_0^{-1}$. 

In its simplest form, the lattice partition function needed
to generate the above random curves is given by
\beq
Z ( \beta ) \; = \; { \cal N } \int \prod_{i=1}^N d^{D} X_i \;
\exp \left \{ - \beta \sum_{i=1}^N | {\bf X}_i - {\bf X}_{i+1} |^{\alpha}
\right \}
\eeq
which for $\alpha=1$ generates closed (${\bf X}_{N+1} = {\bf X}_{1}$)
piecewise linear curves embedded in
$D$ Euclidean dimensions. For $\alpha=2$ it is equivalent to the 
generating function for a one-dimensional $D$-component massless
field theory with unit lattice spacing,
with infrared divergences appearing as the size $N$ goes
to infinity.
In the limit of a large
number of steps $N$ one obtains for the size of the random walk
\beq
< X^2 > \; \equiv \; { 1 \over N } \sum_{i=1}^N < {\bf X}_i^2 >
\; \mathrel{\mathop\sim_{ N \rightarrow \infty }} \; N^{2/d_H}
\eeq
with $d_H = 2$ for free Brownian motion, independently of $\alpha$.
The Hausdorff dimension $d_H$ characterizes the
deviation of an ensemble of random paths from what one would
expect based on their topological dimension of one.
Furthermore $d_H$ is known to be a universal number, i.e. independent 
of the specific choice for the measure over random paths, and describing
general geometric properties of random curves in the limit
of very long paths. 
The relation $\nu^{-1} = d_H$ for free fields follows
as a direct consequence of the representation of the field 
correlation function in terms of a sum over random paths with
fixed endpoints, as given by Eq.~(\ref{eq:corr_rw}).

Below four dimensions non-trivial continuum behavior
is expected for scalar fields, including non-Gaussian exponents
and non-trivial fractal dimensions.
Thus for example for an interacting
scalar field in two dimensions $\nu=1$ (Onsager solution of the Ising model), 
while for a self-avoiding random walk one has instead $d_H=\nu^{-1}=4/3$
exactly in $d=2$ \cite{saw}.
Other non-trivial constraints, such as 
the requirement that the random walks do not back-track, are also expected
to change the fractal dimension $d_H$.

In the gravitational case one is dealing with random paths
associated with a massless particle of spin two. As a result
new constraints on the nature of the random paths
come into play, which are not present in the simpler case
of a spinless scalar field. 
As discussed in Feynman and Hibbs \cite{feypath},
these constraints are in fact already rather complicated
for the case of a particle of spin one-half, and give rise 
even in this simplest case to a set of nontrivial complex weights
needed to correctly reproduce the continuum expression for the
Dirac propagator. 
In four dimensions such paths involve Dirac projection
operators $1 \pm \gamma_\mu$ \cite{lgt}.

On general grounds one would then be inclined to identify the
value $\nu=1/3$ found for four-dimensional gravitation with a fractal
dimension of random gravitational paths $d_H=3$.
Unfortunately (or fortunately) the value $\nu=1/3$ itself does not correspond to any
known field theory or statistical mechanics model in four
dimensions. For dilute branched polymers it is known that $\nu=1/2$
in three dimensions \cite{parsour}, and $\nu=1/4$ at the upper critical
dimension $d=8$ \cite{polymers}, so one would expect a value close to
$1/3$ somewhere in between.
A value for the fractal dimension close to one would indicate
the paths have almost linear Euclidean geometry, while at the opposite
end a very large fractal dimension would indicate the paths
are largely collapsed to a very small region about the origin.
The paths in this latter case are highly folded and to some
extent self-intersecting.
Therefore the value $d_H=3$ found for quantum gravitation would
suggest a far greater degree of folding compared to the spinless case,
for which $d_H=2$. 

One could further develop these arguments and, in analogy with
the scalar case, conclude that below
six space-time dimensions two random gravitational paths will
have a non-vanishing probability of self-intersection.
These arguments would imply that the ``upper critical dimension'' for gravity
is six, above which the theory becomes in some sense non-interacting
and therefore trivial.
\footnote{One can obtain a direct estimate for the upper
critical dimension by equating $ 2 \nu^{-1} (d) = d $, which after
interpolating in $d$ the known Regge lattice results gives as a solution
the surprisingly low value $d \approx 2.929$ (see also Fig.1).
This in turn would lead to the somewhat paradoxical conclusion that quantum gravity,
in spite of being perturbatively non-renormalizable, is weakly interacting
in the infrared above three dimensions.
In the same sense that self-interacting scalar field theories, in spite of
not being perturbatively renormalizable, become weakly interacting at low energies
above four dimensions.}
Unfortunately this argument is probably flawed, as $d_H=3$
holds only in $d=4$, and presumably not at the upper critical dimension,
if one indeed exists.
On the other hand, the large-$d$ geometric estimate discussed previously,
$\nu^{-1} = d-1$, equates to two (the fractal dimension for 
an unconstrained, spinless random walk) in three space-time
dimensions, where it is in fact known that there can be no
propagating, genuine spin two degrees of freedom.

\vskip 30pt
\subsection*{Approaching Quantized Gravitation from Spin Zero and Spin One}
\hspace*{\parindent}

While only limited results exist for the non-perturbative
scaling dimensions of quantum gravitation in four dimensions, the same
is not quite true for spin one (compact Abelian gauge
theory) and spin zero (self-interacting scalar field theory). 
It has been known since the work of Wilson \cite{wilson} that all local
one-component scalar field theories in four dimensions
are described by the Gaussian fixed point. 
The fact that these theories become non-interacting
(up to logarithmic corrections) at large distances in four dimensions
implies that the critical exponents and scaling dimensions
coincide with those of a free field. 
In particular one finds for the universal critical exponent
\beq
\nu^{-1} \; = \; 2 \;\;\;\; (s=0)
\eeq
a result which in fact can be proven rigorously \cite{aizenman}
(for some unresolved issues and
an unconventional point of view regarding the self-interacting
scalar field theory in four dimensions see \cite{klauder}).
In three dimensions the interacting $\lambda \phi^4$ scalar field theory
shares the same universal long distance properties with the 
Ising ferromagnet.
For both incarnations a wealth of numerical and analytical
data exists on
the critical exponents, and for the purposes of the present
discussion the relevant result here is $\nu^{-1} \approx 27/17 = 1.5882$
\cite{zjg}. 

In the massless spin-one case the results for the critical
exponent $\nu$ are somewhat less unambiguous, and furthermore no exact
results are available yet. 
An analytical variational real space renormalization group analysis of the 
Abelian $U(1)$ lattice gauge theory in $d=4$  \cite{u1} gave
\beq
\nu^{-1} \; = \; 2.496(7) \;\;\;\; (s=1) \;\;\; .
\eeq
The errors there can be estimated from an analysis of the results
for $\nu$ using the same renormalization group methods in the 3-d $U(1)$
spin system, which gave $\nu=0.6702(6)$ \cite{u1}, compared to the present
best theoretical value \cite{zjg} $\nu=0.6698(15)$ based on the 
$\epsilon$-expansion about four dimensions as well as the 3-d $\phi^4$ field
theory, and also
in good agreement with the latest experimental value $\nu=0.6706(5)$,
as quoted again in the recent comprehensive review \cite{zjg}.

More recently \cite{lang} detailed
numerical simulation studies of the Abelian $U(1)$ compact lattice gauge theory
have been performed with various additional action terms, besides the standard plaquette term.
Away from the ``Wilson line'' (where the adjoint coupling $\gamma \neq 0$,
and where the transition, being first order, is more difficult to analyze regarding the
true singularity of the free energy located at the end of the metastable phase)
the value $\nu^{-1} = 2.53(6)$ is found for an adjoint
gauge coupling $\gamma=-1/2$.
Closer to the first-order Wilson line $\gamma=0$ they find the 
value $\nu^{-1}=2.74(6)$, but which should really
be discarded in view of the first order nature of the intervening transition
for $\gamma=0$ \cite{schilling}, unless as mentioned above a more refined analytic
continuation towards the true critical point is performed, in order to extract
the required critical exponent characteristic of the end-point singularity.
\footnote{
One might wonder to what extent the quoted numerical results for the critical
exponents, often obtained on relatively small lattices, are reliable.
To further estimate the uncertainties in the numerical determination
for $\nu$ in $d=4$ one can for example compare to a recent high-accuracy
determination of $\nu$ in the spin-zero case (Ising model) in $d=4$,
which yields $\nu^{-1}=1.992(6)$ \cite{parisi_is} on comparable
lattice sizes, and which should be compared to the expected exact
value $\nu^{-1}=2$. }

Given the above values for the critical exponent $\nu^{-1}$ for the massless spin zero
and spin one fields, it is tempting to use them to estimate independently
the scaling dimensions for gravitation, using
\beq
\nu^{-1}(s) \; = \; s \; \nu^{-1}_{s=1} \; + \; (1-s) \; \nu^{-1}_{s=0} \;\; .
\eeq
(see Table I).
With $\nu^{-1}=2$ for $s=0$ and $\nu^{-1}=5/2$ for $s=1$ one then obtains
$\nu^{-1}=3$ for spin two, in good agreement with the previous discussion.
In addition the simple formula
\beq
\nu^{-1} \; = \; 2 + { s \over 2 }
\eeq
gives for the exponent $\alpha / \nu = (2-d \nu)/ \nu =s$ in four dimensions,
and therefore a
divergence of the second derivative of the free energy of the
remarkably simple form $ C \sim \xi^s $.

\begin{table}

\begin{center}
\begin{tabular}{|l|l|l|}
\hline\hline
& & 
\\
Reference & $\nu^{-1}$ in $d=3$ & $\nu^{-1}$ in $d=4$ 
\\ \hline \hline
HW93 \cite{hw3d} & 1.67(6) & 3.34 
\\ \hline
H81 \cite{u1} & - & 2.991
\\ \hline
JLN96 \cite{lang} & -  & 3.05(13)
\\ \hline
CF96 \cite{lang1} & - & 3.48(12)
\\ \hline \hline
exact & 1.5882 & 3 

\\ \hline \hline
\end{tabular}
\end{center}
\label{exp1}

{\small {\it
Table I: Critical Exponent $\nu^{-1}$ for a massless spin two
particle in four dimensions, as 
obtained indirectly either by extrapolation from other dimensions
(d=3 in row 1) or from information on other spin values (rows 2-4).
Included in the table is also one direct (lattice) determination in $d=3$.
The un-weighted average of all extrapolated values listed in the second
column is $\nu^{-1} = 3.22$.
\medskip}}


\end{table}
\vskip 10pt

\begin{table}

\begin{center}
\begin{tabular}{|l|l|l|}
\hline\hline
& & 
\\
Reference & $\nu^{-1}$ in $d=3$ & $\nu^{-1}$ in $d=4$ 
\\ \hline \hline
HW93 \cite{hw3d} & 1.67(6) & -
\\ \hline
H92 \cite{critical} & - & 3.08(62)
\\ \hline
H00 \cite{critical} & - & 2.98(7) 
\\ \hline
AK96 \cite{epsilon} & 1.6 & 4.4 
\\ \hline
RS02 \cite{reuters} & 1.11(5) & 1.68(26)
\\ \hline
RL02 \cite{reuterl} & - & 2.8(6) 
\\ \hline
Li03 \cite{litim} & 1.2 & 2.666
\\ \hline \hline
exact & 1.5882 & 3 

\\ \hline \hline
\end{tabular}
\end{center}
\label{exp2}

{\small {\it
Table II: Direct determinations of the critical exponent $\nu^{-1}$
for quantum gravitation, using a variety of analytical and numerical
methods in three and four space-time dimensions.
The un-weighted average of all direct determinations for quantum
gravitation in four dimensions listed above gives $\nu^{-1}=2.93$.
\medskip}}


\end{table}
\vskip 10pt

Yet another, independent way of estimating the critical exponent for
four-dimensional quantum gravitation involves looking at the two lower
dimensional cases of pure gravity in $d=2$ 
(where $\nu^{-1}=0$) and pure gravity in $d=3$ (where $\nu^{-1} \approx 1.67$)
\cite{hw3d}. A linear extrapolation to four dimensions would then gives
$\nu^{-1} = 3.3$ which is quite consistent with what has been said in the
previous discussion.
It is worth noting here that the value for the exponent for
three-dimensional gravity is tantalizingly close to the scalar field
case. In the $2+ \epsilon$-expansion one finds $\nu^{-1} =1.6$
while some relatively old direct numerical simulations in $d=3$ give $\nu^{-1} =1.67$.
Both values are quite close to the $3-d$ scalar field exponent
$\nu^{-1} =27/17=1.5882$ \cite{zjg}, which would be in line with the 
conjecture that in the infrared limit three-dimensional gravity
belongs to the same universality class as a self-interacting
single-component scalar field, with the scalar curvature
playing the role of the scalar field $ R \sim \Box \phi$, as
in fact suggested some time ago by the authors of reference \cite{deser}.

\begin {figure}
  \begin {center}
    \input{plot1} 
  \end {center}  
\noindent
{\small Fig.\ 1 . Gravitational critical exponent
$\nu^{-1}= - \beta'(G_c)$ as a function of dimension.
Direct determinations from the Regge lattice
(small circles at two, three and four dimensions), in the
continuum using renormalization group truncation
methods (squares), and by extrapolating lattice results from lower spin
(triangles) are compared (see Tables I and II).
The solid line is an interpolation through the Regge lattice results,
incorporating the asymptotic behavior $d-1$ for large $d$.
The thin-dotted line is the analytic $2+\epsilon$ result of
Eq.~(\ref{eq:nueps}).
The dotted line is the continuum renormalization group
result of \cite{litim}.
The origin, methodology and comparison of the these various
results is discussed further in the text.
\medskip}
\end {figure}


As one last exercise one can
look at the case of fractional spin, which presumably corresponds to
massless self-interacting fermions.
In the spin one half case, which should apply to
self-interacting fermions in four dimensions (such as those represented
by the non-renormalizable 4-d Gross-Neveu \cite{gn} and similar four-fermion
models),
one obtains $\nu^{-1} = 9/4 = 2.25$ in $d=4$, which should be compared to the
known values $\nu^{-1} \approx 27/17 =1.5882$ in $d=3$
(interacting 3-d fermions as described by Ising model exponents), and
$\nu = 1$ in $d=2$ (based on the rigorous equivalence between the
two-dimensional critical Ising model and a free Majorana fermion).
Had one extrapolated linearly these known results to four dimensions,
one would have estimated $2.18$, a value quite close to $9/4$
(to within three percent). 
It is of course not obvious at this point how to interpret the above
result in terms of a fermion random walk,
which would have fractal dimension $d_h=\nu^{-1}=9/4$.
But the trend in the exponents is at least consistent with
the expectation that the fractal dimension increases 
with embedding dimension $d$, as there are more dimensions 
to expand into.

\begin{table}

\begin{center}
\begin{tabular}{|l|l|l|l|}

\hline\hline
& & &
\\
$\nu^{-1}$ & $d=2$ & $d=3$ & $d=4$ 
\\ \hline \hline
spin s=0 & 1 & 1.588 & 2 
\\ \hline
spin s=1 & 0 & 0     & 2.5 
\\ \hline
spin s=2 & 0 & 1.588  & 3 
\\ \hline \hline

\end{tabular}
\end{center}
\label{exp3}

\center{\small {\it
Table III: Summary table for the critical exponent $\nu^{-1}$ as
a function of spin and dimension.
\medskip}}

\end{table}

\vskip 10pt

The various estimates for the critical exponent are compared
in Tables I (indirect determinations) and II (direct determinations).
Table II provides a list of critical
exponents for gravitation as obtained by direct perturbative
and non-perturbative methods in three and four dimensions.
As mentioned before, direct numerical simulations
for the lattice model of Eq.~(\ref{eq:zlatt}) in four dimensions
give for the critical point $G_c=0.626(11)$ in units of
the ultraviolet cutoff, and one finds \cite{critical}
\beq
\nu^{-1} \; = \; 2.99(8) \;\;\;\; (s=2)
\eeq
which is used for comparison in Table II.
The fact that the critical coupling $G_c$ is not small shows incidentally that
the lattice theory is not weakly coupled close to
the transition point.
\footnote{
Furthermore, the critical point obtained from the
analytic continuation of the strong coupling (small $k$) branch
of the free energy lies at the end of the metastable phase
of the Euclidean theory, which is not necessarily a concern here
as one is ultimately interested in the pseudo-Riemanniann 
theory. Indeed one would not have expected otherwise, in view of the 
well-known and seemingly un-avoidable conformal instability of the
Euclidean theory.}

To conclude this section, one can reverse the line of the above arguments
relating to the critical exponents for gravitation, and
instead estimate the spin of the massless lattice graviton by considering
the dependence of the measured exponent $\nu$ on the spin.
Assuming a linear dependence of the exponent $\nu^{-1}$ on
the spin and using the most accurate values at $s=0$ and $s=1$ one
obtains, from $\nu^{-1}=2.98(7)$ \cite{critical}, about
$s=1.98(3)$, which is quite close to the expected value
of spin two. 

\newpage

\vskip 30pt
\newsection{Exponents and Long-Distance Quantized Gravitation}
\hspace*{\parindent}

The result of Eq.~(\ref{eq:grun1}) implies that the gravitational
constant is no longer constant as in the classical theory,
but instead slowly changes with scale due to the presence
of weak vacuum polarization effects,
\beq
G(r) \; = \; G(0) \left [ \; 1 \, + \, c \, ( r / \xi )^{1 / \nu} \, 
+ \, O (( r / \xi )^{2 / \nu} ) \; \right ] \;\; ,
\label{eq:grun2}
\eeq
The exponent $\nu$, related to the derivative
of the beta function evaluated at the non-trivial ultraviolet
fixed point via the relation $ \beta ' (G_c) \, = \, - 1/ \nu = -3 $
(see the previous discussion in Section 4 and reference \cite{critical})
for pure quantum gravitation,
is supposed to characterize the universal long-distance properties of quantum
gravitation, and is therefore expected to be independent of the specifics
related to the nature of the ultraviolet regulator, introduced to make
the quantum theory well defined.

The mass scale $m = \xi^{-1}$ in Eq.~(\ref{eq:grun2}) determines the
magnitude of quantum deviations from the classical theory, and separates the short distance,
ultraviolet regime with characteristic momentum scale $ \mu \ll m $ where
non-perturbative quantum corrections are negligible, from
the long distance regime where quantum corrections are significant.
It should be emphasized here that most of these considerations are in fact
quite general, to the extent that they rely mainly on rather general principles of
the renormalization group and are in fact not tied to any particular value for
the exponent $\nu$, although the value $\nu=1/3$ clearly has some aesthetic appeal. 
Furthermore the dimensionless constant $c$ is, at least in principle, a calculable number.
In \cite{corr} $c$ was estimated from the curvature
correlation function at $c=0.014(4)$, while more
recently in \cite{lines} it was estimated to be $c=0.056(27)$
from the correlation of Wilson lines.
It is important to note that while the exponent $\nu$ is universal,
$c$ in general depends on the specific choice of regularization scheme
(i.e. lattice regularization versus dimensional regularization or
momentum subtraction scheme).

It is worthwhile to note that the result of Eq.~(\ref{eq:grun2}),
which as discussed in Section 4 is
a direct consequence of Eq.~(\ref{eq:chising}) and the value for
$\nu$ found in the lattice theory (defined by the partition function
of Eq.~(\ref{eq:zlatt}) with higher derivative
coupling ($a \rightarrow 0$) and functional measure parameter $\sigma=0$),
only applies to the simplest case of pure Einstein gravity with a bare
cosmological term.
\footnote{Light matter fields will modify the exponent $\nu$, and
therefore the result of Eq.~(\ref{eq:grun2}), provided their
mass is small enough to contribute significantly to
vacuum polarization loops, $m \sim \xi^{-1}$.}
But one does not expect this to be the correct theory at sufficiently
short distances $r \sim l_P$, where higher derivative curvature terms
will come into play, either through direct inclusion and gravitational
radiative corrections, or via matter field and the conformal anomaly. 
In this limit the gravitational potential will be further modified
by exponential and logarithmic terms, as discussed in reference \cite{hdqg}.
 
Let us recall here that in $SU(N)$ gauge theories and in particular in QCD,
the theory of the strong interactions, a similar set
of results is known to hold \cite{frampton}.
The crucial difference lies in the fact that there the
scale evolution of the coupling constant can be systematically computed
in perturbation theory due to asymptotic freedom, a 
statement which reflects the fact that such theories become
free at short distances (up to logarithmic corrections).
In non-Abelian gauge theories one has for weak coupling
theory 
\beq
{ 1 \over g^2 (\mu) } \; = \; { 1 \over g^2 ( \Lambda_{\overline{MS}} ) } 
\; + \; 2 \beta_0 \; \log \left (
{ \mu \over \Lambda_{\overline{MS}} } \right ) \; + \cdots
\label{eq:qcdrun}
\eeq
with $\beta_0 = { 1 \over 16 \pi^2 } ( {11 N \over 3} - {2 \over 3} n_f )$
where $N$ is the number of colors, $n_f$ is the number of flavors
of massless fermions, $\mu=1/r$ is an arbitrary momentum scale, and
$\Lambda_{\overline{MS}}=200 MeV$ is a scale parameter which determines
the size of scaling violations.  The $\cdots$'s indicate
higher order loop corrections.
Of course QCD does not determine $\Lambda_{\overline{MS}}$ (it appears as an
integration constant of the Callan-Symanzik renormalization group equations),
and therefore it has to be fixed by experiment from a measurement
of the size of scaling violations, i.e. via the observed deviations
from free field behavior at sufficiently high energies. 
It is a remarkable fact that a good fraction of QCD and
electroweak standard model phenomenology simply follows
from the result in Eq.~(\ref{eq:qcdrun}) and its electroweak
analog. 
\footnote{
In QED the scale dependence of the vacuum polarization effects is
of course quite small, with the fine structure
constant only changing from $\alpha (0) \sim 1/137.036$ at atomic distances
to about $ \alpha (m_{Z_0}) \sim 1/128.978 $ at energies comparable to
the $Z^0$ mass.}

If one pursues in a straightforward way
the analogy with non-Abelian gauge theories
one is led to conclude that in
quantum gravitation the quantity $\xi$ plays the same role
as $\Lambda_{\overline{MS}}$ in QCD, 
$\xi \leftrightarrow \Lambda_{\overline{MS}}$.
One major difference between the two theories lies of course in the
fact that in one case the ultraviolet fixed point is at the
origin $g^2=0$, while in the other it is not.
As a result one has logarithmic quantum corrections to free field
behavior in QCD, but power law corrections in gravitation.

To determine the actual physical value for the non-perturbative
scale $\xi$ further physical input is needed.
It seems natural to identify  $ 1 / \xi^2 $ with either some average
spatial curvature, or perhaps more appropriately with the Hubble constant
determining the macroscopic expansion rate of the present universe
\cite{det,critical}, via the correspondence
\beq
\xi \; = \; 1 / H_0 \;\; ,
\label{eq:hub}
\eeq
in a system of units for which the speed of light is equal to one.
This correspondence can be elaborated upon further.
In the standard homogeneous isotropic Friedmann-Robertson-Walker model of classical 
relativistic cosmology one uses the line element
\beq
ds^2 \; = \; dt^2 \; - \; R^2 (t) \left \{ {d r^2 \over 1 - k r^2 } 
\; + r^2 d \theta^2 + r^2 \sin^2 \theta \; d \phi^2 \right \}
\eeq
with $k=0, \pm 1$ and $H (t_0) = ( {\dot R} / R )_{t_0} $ denoting 
today's expansion rate as determined from the field equations.
It is well known that the presence of a small cosmological constant induces 
an exponential expansion of the scale factor at large times.
In this very distant future, dominated by a non-vanishing cosmological
constant, an equivalent description can be given
in terms of the static isotropic de Sitter metric 
\beq
ds^2 \; = \; \left ( 1 - H^2_\infty r^2 \right ) dt^2 \; - \;
\left ( 1 - H^2_\infty r^2 \right )^{-1} dr^2 
\; - \; r^2 ( d \theta^2 + \sin^2 \theta d \phi^2 )
\eeq
with a horizon radius $H_\infty = \lim_{t \rightarrow \infty} H(t)$.
From Einstein's classical field equations one has
\beq
H^2_\infty \; = \; { 8 \pi G \over 3 } \lambda \; \equiv \; { \Lambda \over 3 }
\eeq
so the existence of an $H_\infty$ is equivalent to assuming the
presence of a non-vanishing cosmological constant $\lambda$
(here we follow common convention in defining the scaled
cosmological constant $\Lambda$, which
should not be confused with the ultraviolet cutoff).
It is presumably this quantity which should be identified with $\xi$.
Given the rather crude nature of our arguments, in the following we shall
not distinguish between $H_0$ and $H_\infty$,
and simply take $H_0^{-1} \sim 10^{28} cm $ as today's
estimate for the size of the visible universe. 
\footnote{
While the observational evidence
for a non-vanishing cosmological constant is quite recent,
simplicial lattice theories of Euclidean quantum gravity can, as far as
one knows, only be formulated with a {\it non-vanishing} positive
bare cosmological constant ($\lambda > 0$). 
In the absence of such a constant the path integral
does not converge for large edge lengths, and {\it no} stable ground state
exists in the Euclidean theory \cite{hw84,lesh,monte,hartle1}. }

The appearance of the renormalization group invariant
quantity $\xi$ in the quantum evolution of the coupling
$G$, a very large quantity by the identification of
Eq.~(\ref{eq:hub}), suggests that the leading
scale-dependent correction, which gradually increases the strength of the
effective gravitational interaction as one goes to larger and
larger length scales, should be extremely small.
One would therefore expect the deviations from classical general
relativistic behavior for most physical quantities to be in the end
practically negligible, at least until one reaches 
very large distances $r \sim \xi$.

At this stage we should comment on an apparent paradox associated
with the identification of the correlation length $\xi$ with
$1 / H_0$.
Naively one would expect, simply on the basis of dimensional
arguments, that the curvature scale close to the fixed point 
be determined by the correlation length
\footnote{There is at least in principle an even more naive expectation,
namely ${\cal R} \sim 1/l_P^2$, which is excluded though by all
numerical studies of simplicial lattice gravity.}
\beq
{\cal R} \; \mathrel{\mathop\sim_{ {\cal R} \rightarrow 0}} \; 1/ \xi^2 \;\; ,
\label{eq:naive}
\eeq
but one cannot in general exclude the appearance of some non-trivial exponent.
Indeed one finds for the vacuum expectation value of
the Ricci scalar (see Eqs.~(\ref{eq:rk}) and (\ref{eq:m_latt}))
\beq
{\cal R} ( \xi ) \; \mathrel{\mathop\sim_{ k \rightarrow k_c}} \;
{ 1 \over l_P^{2-d+1/\nu} \xi^{d-1/\nu} }
\label{eq:rm1}
\eeq
with $\nu=1/3$ in four dimensions, and therefore $ {\cal R} \sim 1 / l_P \; \xi $.
Only close to two dimensions one recovers the classical
result, for which $\nu \sim 1/(d-2)$.

At first one might be tempted to identify the expectation value
of the local scalar curvature with the quantity $ H_0^2 $,
but further thought reveals that this correspondence is inconsistent
with the identification $\xi= 1 / H_0$ proposed before, and would
only be legitimate if the local curvature average
${\cal R}$ were to indeed correctly describe the rotation
of vectors, parallel transported around very large loops.
But the analogy with the local action density $< F_{\mu\nu}^2 >$ in non-Abelian
gauge theories seems to suggest that such an identification
might in fact be incorrect \cite{smit}, and that the long-distance contribution
to the curvature is not given by the local average in Eq.~(\ref{eq:rm1}), but
should instead be measured directly by computing
the parallel transport of vectors around very large loops
(with characteristic size much larger that the Planck length,
$A \sim \xi^2 \gg l_p^2 $), but this is laborious and has not been done yet.
Indeed at the other end one expects, for very short distances
comparable to the size of the ultraviolet cutoff,
significant fluctuations in the curvature
with fluctuations of the order of ${\cal R} \sim 1/ l_P^2 $.
In other words, the above arguments would suggest that the observable 
average curvature should be scale dependent.

On the other hand for the curvature correlation at fixed geodesic distance $d$
one obtains from simple scaling, and for 
``short distances'' ($r \ll \xi$),
\beq
< \sqrt{g} \; R(x) \; \sqrt{g} \; R(y) \; \delta ( | x - y | -d ) >_c \;
\mathrel{\mathop\sim_{d \; \ll \; \xi }} \;\; 
{1 \over d^{\; 2 \; (d-1/ \nu)} } \;\; \sim \;\; {A \over d^2 } \;\;\; ,
\label{eq:pow2}
\eeq
for $\nu=1/3$ in four dimensions,
and with $A$ a calculable constant of order one \cite{critical}.
This last result follows almost immediately from the relationship between the
curvature-curvature correlation function and the second derivative of the
partition function with respect to $G$, which determines the curvature
fluctuation and thus the curvature correlation function at zero momentum. 
If one then considers the (scalar) curvature $R$ averaged over a very
small spherical volume $V_r = 4 \pi r^3 /3$,
\beq
\overline{ \sqrt{g} \; R } \; = \; { 1 \over V_r } \;
\int_{V_r} d^3 {\bf x} \;  \sqrt{g({\bf x},t)} \; R({\bf x},t)
\eeq
one can compute the corresponding variance as
\beq
\left [ \delta ( \sqrt{g} \; R ) \right ]^2 \; = \; { 1 \over V_r^2 } \;
\int_{V_r} d^3 {\bf x} \; \int_{V_r} d^3 {\bf y} \;
< \sqrt{g} \; R({\bf x}) \; \sqrt{g} \; R({\bf y}) >_c \; = \; 
{9 A \over 4 \; r^2 } \;\;\; .
\eeq
As a result the r.m.s. fluctuation of $\sqrt{g} R$ averaged over a 
small spherical region of size $r$ is given by
\beq
\delta ( \sqrt{g} \; R ) \; = \; 
{ 3 \sqrt{A} \over 2 } \; {1 \over r } \;\;\; ,
\eeq
with a Fourier transform power spectrum of the form
\beq
P_{\bf k} \; = \; \vert \; \sqrt{g} \; R_{\bf k} \; \vert^2 \; = \; 
{ 4 \pi^2 A \over 2 V } \; {1 \over k } \;\;\; .
\eeq
These results only hold for relatively short distances, and presumably get
modified at distances $r \sim \xi$.
The semi-classical answer would look quite different;
in this limit $\nu \sim 1/(d-2)$ and therefore for the curvature
correlation the distance dependence would tend to $1/d^4$ close to $d=2$,
where it makes sense to make a comparison.

One can go one step further and
use Einstein's field equations to relate the local curvature to the
local mass density. From the field equations
\beq
R_{\mu\nu} - \half \, g_{\mu\nu} R \; = \; 8 \pi G \; T_{\mu\nu} \;\;
\label{eq:ein}
\eeq
for a perfect fluid
\beq
T_{\mu\nu} \; = \; p \, g_{\mu\nu} + (p+\rho) \, u_\mu u_\nu \;\;
\label{eq:fluid}
\eeq
one obtains for the Ricci scalar, in the limit of negligible pressure,
\beq
R (x) \; \approx \; 8 \pi G \; \rho (x) \;\;\; .
\eeq
As a result one obtains from Eq.~(\ref{eq:pow2}) for the density
fluctuations a power law decay of the form
\beq
< \rho (x) \; \rho (y) >_c \;
\mathrel{\mathop\sim_{ |x-y| \; \ll \; \xi }} \;
{1 \over |x-y|^2 } \;\; .
\label{eq:pow3}
\eeq
One can list a few other classical general relativistic results
which are presumably affected by a running gravitational constant.
It should be clear from the
above discussion that in
order for the quantum corrections to become quantitatively significant,
one needs to look at rather large distance scales comparable to $\xi$, or in
other words $r \sim 1 / H_0$.
In standard classical cosmology one writes
\beq
H_0^2 \; = \; { 8 \pi G(r) \over 3 } \left [ 
\; \rho_\Lambda + \rho_{DM}  + \rho_B \; \right ]
\label{eq:density}
\eeq
with $G$ usually assumed to be constant, and with
the $\rho$'s representing various density contributions.
On the l.h.s one usually neglects terms of order
$k / R_0^2 $ arising from the curvature of the hypersurface
of homogeneity.
In view of the what has been said before though it
seems natural that $G(r)$ in the above expression should be taken
at the largest length scale $ H_0^{-1}$. 
Then one obtains for the overall coefficient
a quantity slightly larger than the laboratory value $ \sim G(0)$, namely
$G(H_0^{-1}) \approx G(0)\; (1+c) > G(0) $.
On the lattice one finds a rather small value for $c \approx 0.06$.
One should recall however, as stated earlier, that while the exponent
$\nu$ is universal, the quantity $c$ is not, and in general depends on the specific
regularization scheme. 
More specifically, in ordinary lattice gauge theories one finds large but calculable
finite renormalization factors, relating the lattice gauge coupling to
the continuum coupling \cite{hasen}. 
A more reasonable expectation would therefore be that $G(H_0^{-1})$ is
related to $G(0)$ by a constant of proportionality which is roughly
of order one.
Additional cosmological and astrophysical arguments and proposed tests
can be found in \cite{vipul}.

\newpage

\vskip 30pt
\section{Concluding Remarks}
\hspace*{\parindent}

In this paper we have examined some aspects of the
connection between lattice and continuum models for quantum 
gravity. 
In particular the aim of the paper was to elucidate the
relationship between the more recent simplicial lattice results,
which do not rely on the weak field expansion and
are therefore inherently non-perturbative,
and the semiclassical Euclidean functional integral expansion in the continuum.

The first issue addressed was the very definition of the
notion of spin content in the lattice theory. 
Proceeding from the Euclidean Feynman path integral approach,
we have constructed the lattice analog of the semiclassical expansion
for the ground state functional of linearized gravity.
Two procedures were followed.
The first procedure relied on constructing directly a
lattice expression for the exponent of the ground state functional,
obtained by transcribing the continuum expression
in terms of lattice variables. 
There one proceeds from
the lattice expression for the gravitational action, specified
on a fixed time slice, and supplemented by the appropriate vacuum
gauge conditions.
A crucial ingredient in this method is the correct identification
of the correspondence between continuum degrees of freedom
(the metric) and the lattice variables (the squared edge lengths).
The resulting lattice expression is then equivalent to the continuum
one by construction.

The second procedure relies instead only on the expression for the
lattice gravitational action, as expanded in the weak field limit,
and determines the explicit lattice
form for the ground state functional for linearized gravity
by performing explicitly the necessary lattice Gaussian functional integrals.
The resulting discrete expression can then be compared to
the continuum one by systematically re-expressing the edge lengths in terms
of the metric. 
It is encouraging that the resulting lattice expression completely agrees
with what is found by using the previous method.

It is advantageous in performing the above calculation to
introduce spin projection operators, which separate out
the spin zero, spin one and spin two components of the
gravitational action.
As a by-product one can show that the lattice gravitational action 
only propagates massless spin two (or transverse-traceless)
degrees of freedom in the weak field limit, as is the case in the continuum. 
Furthermore, as expected the lattice ground state functional
for linearized gravity only contains these physical modes.

The explicit construction of the ground state wave functional
of linearized lattice gravity in terms of the lattice transverse-traceless
modes makes it possible at least in principle to compare
the lattice and continuum results in the limit of small
curvatures.
After imposing appropriate boundary conditions at infinity by suitably restricting
the values for the edge lengths on the lattice
boundary such that the deficit angle is zero there,
one would then have to enforce as well the lattice vacuum
gauge conditions of Eq.~(\ref{eq:gauge}) so as to make contact
with the semiclassical lattice functional of Eq.~(\ref{eq:ttact}).
Since no gauge fixing is required for determining invariant averages obtained via
the partition function of Eq.~(\ref{eq:zlatt}), 
the gauge conditions would have to be imposed 
on each edge length configuration, by progressively
applying local gauge transformations \cite{gauge}.
But it is expected that after such a transformation the edge distributions
on a fixed time slice should follow closely the distribution
of Eq.~(\ref{eq:ttact}), if indeed as expected the only surviving physical modes
are transverse traceless.

In subsequent sections of the paper we have systematically
examined the relationship
between recent non-perturbative results obtained in the lattice theory
and the corresponding calculations as performed in the continuum theory.
The latter suggest the presence of a non-trivial ultraviolet
fixed point in $G$, and in some cases have led to definite
predictions for the universal critical exponent of quantum
gravitation, which can therefore be compared - even quantitatively -
to the lattice results.

Besides relying on the recent lattice and continuum results
for quantum gravitation, one can also independently try
to estimate the gravitational scaling dimensions based on
what is known based on exact and approximate
renormalization group methods for spin zero 
(self-interacting scalar field in four dimensions)
and spin one (Abelian non-compact gauge theories), for which
a wealth of information is available on the critical indices.
We have argued that these results too are remarkably consistent
with what is known about the gravitational exponents in four dimensions,
both from the discrete as well as from the continuum side.
We have also presented a simple geometric argument
which interprets the value for the gravitational exponent $\nu^{-1}=3$.

In the last section of the paper we have discussed some (almost immediate)
physical implications of recent lattice and continuum results,
with an emphasis on the small expected deviations from
classical general relativity expected at sufficiently large scales due
to the running of $G$.
We have argued that it is an almost inevitable consequence
of the existence of an ultraviolet fixed point that the
gravitational coupling becomes scale dependent, with power
law corrections involving the correlation length.
In analogy with non-Abelian gauge theories, and in the
absence of any other likely physical candidate,
it seem natural to identify the non-perturbative scale determining
the size of deviations from classical gravitation with $1/H_0$,
as suggested in \cite{det}.

\vspace{20pt}

{\bf Acknowledgements}

The authors wish to thank James Hartle for suggesting we look at the
formulation of transverse-traceless modes on the lattice and for
suggesting a Regge calculus version of the continuum semiclassical
expansion for the ground state functional, and for many helpful discussions.
The work of Ruth Williams was supported in part by the UK Particle
Physics and Astronomy Research Council.

\vfill

\newpage


\vspace{20pt}


\newpage



\begin{thebibliography}{99}

\bibitem {fey}
R.~P.~Feynman, {\it Act.\ Phys.\ Pol.} {\bf 24} 697 (1963); \\
B.~DeWitt, {\it Phys.\ Rev.} {\bf 160} 1113 (1967);
{\it Phys.\ Rev.} {\bf 162} 1195, 1239 (1967).

\bibitem {hooft}
G.~ 't Hooft and M.~Veltman, {\it Ann. Inst. Poincar\'e} {\bf 20}
69 (1974); \\
S.~Deser and P.~van Nieuwenhuizen, {\it Phys. Rev.} {\bf D10}
401, 410 (1974)

\bibitem{wilson}
K.~G.~Wilson, {\it Rev.\ Mod.\ Phys.} {\bf 47} 773 (1975); {\bf 55} 583 (1983);
{\it Phys.\ Rev.} {\bf D7} 2911 (1973); \\
K.~G.~Wilson and M.~Fisher, {\it Phys.\ Rev.\ Lett.} {\bf 28} 240 (1972).

\bibitem {parisi}
G.~Parisi, {\it Nucl.\ Phys.} {\bf B100} 368 (1975), {\it ibid.} 
{\bf 254} 58 (1985); {\sl `On NonRenormalizable Interactions'}, 
in the proceedings of the Cargese Summer Institute (1976).

\bibitem{nonren}
K.~Gawedzki and A.~Kupiainen, 
{\it Phys.\ Rev.\ Lett.} {\bf 54} 2191 (1985); {\bf 55} 363 (1985);
{\it Nucl.\ Phys.} {\bf B262} 33 (1985);
C.~de Calan, P.~A.~Faria da Veiga, J.~Magnen, R.~Seneor, 
{\it Phys.\ Rev.\ Lett.} {\bf 66} 3233 (1991); \\
S.~Weinberg, {\it Phys.\ Rev.} {\bf D56} 2303 (1997).

\bibitem {regge}
T.~Regge, {\it Nuovo Cimento} {\bf 19} 558 (1961).

\bibitem {rowi}
M.~Ro\u cek and R.~M.~Williams, {\it Phys.\ Lett.} {\bf 104B} 31 (1981);
{\it Z.\ Phys.} {\bf C21} 371 (1984).

\bibitem {hw84}
H.~W.~Hamber and R.~M.~Williams, {\it Nucl.\ Phys.} {\bf B248} 392 (1984);
{\bf B260} 747 (1985); 
{\it Phys.\ Lett.} {\bf 157B} 368 (1985);
{\it Nucl.\ Phys.} {\bf B267} 482 (1986); {\bf B269} 712 (1986).

\bibitem{lesh} H.~W.~Hamber, {\sl `Simplicial Quantum Gravity'},
in {\it Critical Phenomena, Random Systems, Gauge Theories}, 
{\sl 1984 Les Houches Summer School}, Session XLIII, edited by
K.~Osterwalder and R.~Stora (North Holland, Amsterdam, 1986);
{\it Nucl.\ Phys.\ Proc.\ Suppl.} {\bf 25A} 150-175 (1992).

\bibitem {critical}
H.~W.~Hamber, {\it Phys.\ Rev.} {\bf D61} 124008 (2000);
{\it Phys.\ Rev.} {\bf D45} 507 (1992).

\bibitem {dewittm}
B.~DeWitt, {\it Phys. Rev.} {\bf 160} 1113 (1967);
\\
K.~Fujikawa, {\it Nucl. Phys.} {\bf B226} 437 (1983).
 
\bibitem {misnerm}
C.~W.~Misner, {\it Rev. Mod. Phys.} {\bf 29} 497 (1957);
\\
L.~Fadeev and V.~Popov, {\it Sov. Phys. Usp.} {\bf 16} 777 (1974);
{\it Usp.\ Fiz.\ Nauk.} {\bf 109} 427 (1974).

\bibitem {hartle1}
J.~B.~Hartle, {\it J.\ Math.\ Phys.} {\bf 26} 804 (1985);
{\bf 27} 287 (1985); {\bf 30} 452 (1989).

\bibitem {gauge}
H.~W.~Hamber and R.~M.~Williams, {\it Nucl.\ Phys.} {\bf B487} 345 (1997);
{\bf B451} 305 (1995).

\bibitem {det}
H.~W.~Hamber and R.~M.~Williams, {\it Phys.\ Rev.} {\bf D59} 064014 (1999).

\bibitem {monte}
B.~Berg,  {\it Phys.\ Rev.\ Lett.} {\bf 55} 904 (1985);
{\it Phys.\ Lett.} {\bf B176} 39 (1986).

\bibitem {monte1}
W.~Beirl, E.~Gerstenmayer, H.~Markum and J.~Riedler,
{\it Phys.\ Rev.} {\bf D49} 5231 (1994);
J.~Riedler, W.~Beirl, E.~Bittner, A.~Hauke, P.~Homolka and H.~Markum,
Class.\ Quant.\ Grav.\  {\bf 16}, 1163 (1999);
E.~Bittner, W.~Janke and H.~Markum, {\it Phys.\ Rev.} {\bf D66}, 024008 (2002).

\bibitem {frampton}
P.~Frampton, {\sl `Gauge Field Theories'}, (Wiley, New York 2000), 
ch. 6 and references therein.

\bibitem{dtrs}
J.~Ambjorn, M.~Carfora and A.~Marzuoli,
{\sl The Geometry of Dynamical Triangulations}, Lecture Notes
in Physics (Springer Verlag, Berlin, 1997), and references therein.

\bibitem{kuchar} K.~Kucha\u r, {\jmp} {\bf 11}, 3322 (1970).

\bibitem{hartle} J.~B.~Hartle, {\pr} {\bf 29}, 2730 (1984).

\bibitem {hawking}
S.~W.~Hawking, in {\sl `General Relativity - An Einstein Centenary
Survey'}, edited by S.~W.~Hawking and W.~Israel,
(Cambridge University Press, 1979).

\bibitem {veltman}
M.~Veltman, in {\sl `Methods in Field Theory'},
Les Houches Lecture notes, Session XXVIII (North Holland, 1975).

\bibitem{vnh}
P.~van Nieuwenhuizen, {\it Nucl.\ Phys.} {\bf B60} 478 (1973).

\bibitem{rocek1} M.~Ro\u cek and R.~M.~Williams, {\sl `Introduction to
quantum Regge calculus'}, in {\it Quantum Structure of Space and Time},
edited by M.~J.~Duff and C.~J.~Isham (Cambridge University Press,
1982).

\bibitem {hw3d}
H.~W.~Hamber and R.~M.~Williams, {\it Phys.\ Rev.} {\bf D47} 510 (1993).

\bibitem{wheeler} J.~A.~Wheeler, {\sl `Regge calculus and
Schwarzschild geometry'}, in  {\it Relativity, Groups and
Topology}, edited by B.~DeWitt and C.~DeWitt (Gordon and Breach, New
York, 1964).

\bibitem{gross}
D.~Gross and F.~Wilczek, {\it Phys.\ Rev.\ Lett.} {\bf 30} 1343 (1973); \\
H.~D.~Politzer {\it Phys.\ Rev.\ Lett.} {\bf 30} 1346 (1973); see also \\
G.~'t Hooft, {\it Nucl.\ Phys.} {\bf B254} 11 (1985).

\bibitem {cms}
J.~Cheeger, W.~M\"uller and R.~Schrader, {\sl Comm.\ Math.\ Phys.} {\bf 92}
405 (1984); 
and in {\sl `Unified Theories Of Elementary Particles'},  
{\it Heisenberg Symposium}, (Springer, New York, 1982).

\bibitem {lee}
T.~D.~Lee, in {\sl `Discrete Mechanics'}, 
1983 Erice International School of Subnuclear Physics,
vol. 21 (Plenum Press, New York 1985); 
R.~Friedberg and T.~D.~Lee, {\it Nucl. Phys.} {\bf B242} 145 (1984).

\bibitem {lines}
H.~W.~Hamber and R.~M.~Williams, {\it Nucl.\ Phys.} {\bf B435} 361 (1995).

\bibitem {epsilon}
T.~Aida and Y.~Kitazawa, {\it Nucl.\ Phys.} {\bf B491} 427 (1997).

\bibitem {epsilon1}
R.~Gastmans, R.~Kallosh and C.~Truffin, {\it Nucl.\ Phys.} {\bf B133} 417 (1978); \\
S.M. Christensen and M.J. Duff, {\it Phys.\ Lett.} {\bf B79} 213 (1978).

\bibitem {epsilon2}
H.~Kawai and M.~Ninomiya,  {\it Nucl.\ Phys.} {\bf B336} 115 (1990); \\
H.~Kawai, Y.~Kitazawa and M.~Ninomiya,  {\it Nucl.\ Phys.} {\bf B393} 280 (1993) and {\bf B404} 684 (1993); \\
T.~Aida, Y.~Kitazawa, J.~Nishimura and A.~Tsuchiya, {\it Nucl.\ Phys.} {\bf B444} 353 (1995); \\
Y.~Kitazawa, {\it Nucl.\ Phys.} {\bf B453} 477 (1995); \\
Y.~Kitazawa and M.~Ninomiya, {\it Phys.\ Rev.} {\bf D55} 2076 (1997).

\bibitem{epsilon3}
S.~Weinberg, in {\sl `General Relativity - An Einstein Centenary
Survey'}, edited by S.W. Hawking and W. Israel,
(Cambridge University Press, 1979).

\bibitem {zj}
J.~C.~Le Guillou and J.~Zinn-Justin, {\it Phys.\ Rev.} {\bf B21} 3976 (1980);
and {\it J.\ Phys.} (France) {\bf 50} 1395 (1989);
\\
J.~Zinn-Justin, in {\sl 6th International Conference on Path Integrals},
(PI 98) Florence, Italy (1998).

\bibitem{zjg}
R.~Guida and J.~Zinn-Justin, Saclay Preprint SPhTh-t97/040,
cond-mat/9803240, {\it Nucl.\ Phys.} {\bf B489} 626 (1997),
and references therein.

\bibitem {pbook}
G.~Parisi, {\sl `Statistical Field Theory'}, ch. 10,
Addison Wesley, New York (1988), and references therein.

\bibitem{izbook}
C.~Itzykson and J.~M.~Drouffe, {\sl Statistical Field Theory},
Cambridge University Press (Cambridge, 1989).

\bibitem{zjbook}
J.~Zinn-Justin, {\sl Quantum Field Theory and Critical Phenomena},
Clarendon Press (Oxford, 1989).

\bibitem {brezin}
E.~Brezin and J.~Zinn-Justin, {\it Phys.\ Rev.\ Lett.} {\bf 36} 691 (1976); 
{\it Phys.\ Rev.} {\bf D14} 2615 (1976); \\
E.~Brezin and S.~Hikami, LPTENS-96-64 (Dec 1996); cond-mat/9612016.

\bibitem{gn}
D.~J.~Gross and A.~Neveu, {\it Phys.\ Rev.} {\bf D10} 3235 (1974).

\bibitem{donoghue}
J.~F.~Donoghue, {\it Phys.\ Rev.} {\bf D50} 3874 (1994);
{\it Phys.\ Rev.\ Lett.} {\bf 72} 2996 (1994); \\
H.~W.~Hamber and S.~Liu, {\it Phys.\ Lett.} {\bf B357} 51 (1995); \\
I.~J.~Muzinich and S.~Vokos, {\it Phys.\ Rev.} {\bf D52} 3472 (1995); \\
N.~E.~J.~Bjerrum-Bohr, J.~F.~Donoghue and B.~R.~Holstein,
{\it Phys.\ Rev.} {\bf D67} 084033 (2003), and references therein.

\bibitem{reuters}
M.~Reuter and F.~Saueressig, {\it Phys.\ Rev.} {\bf D65} 065016 (2002);
see also V.~Branchina, K.~A.~Meissner and G.~Veneziano,
{\it Phys.\ Lett.} {\bf B 574}, 319 (2003).

\bibitem{reuterl}
O.~Lausher and M.~Reuter, {\it Class.\ Quant.\ Grav.} {\bf 19} 483 (2002).

\bibitem{litim}
D.~Litim, CERN-TH-2003-299, {\it Phys.\ Rev.\ Lett.} {\bf 92} 201301 (2004). 

\bibitem{lense}
J.~Lense and H.~Thirring, {\it Physik.\ Zeitsch.} {\bf 19} 156 (1918).
Press, 1989).

\bibitem{polybook}
A.~M.~Polyakov, {\sl Gauge Fields and Strings}, ch. 9 (Oxford University
Press, 1989).

\bibitem {mandel}
B.~Mandelbrot, {\sl `The Fractal Geometry of Nature'},
(W. H. Freeman and Co., 1982).

\bibitem{aizenman}
M.~Aizenman, {\it Phys.\ Rev.\ Lett.} {\bf 47} 1 (1981); 
{\it Commun.\ Math.\ Phys.} {\bf 86} 1 (1982); {\bf 97} 91 (1985); \\
J.~Fr\"ohlich, {\it Nucl.\ Phys.} {\bf B200} 281 (1982).

\bibitem{luescher}
M.~L\"uscher,  {\sl `Solution of the Lattice $\phi^4$ Theory 
in 4 Dimensions'},
in {\sl Non-Perturbative Quantum Field Theory}, Cargese 1987.

\bibitem{klauder}
J.~R.~Klauder, {\it Lett.\ Math.\ Phys.} {\bf 63}, 229 (2003);
and preprint hep-th/0306016.

\bibitem {saw}
J.~Cardy and H.~W.~Hamber, {\it Phys.\ Rev.\ Lett.} {\bf 45} 499 (1980); \\
B.~Nienhuis, {\it Phys.\ Rev.\ Lett.} {\bf 49} 1063 (1982).

\bibitem {feypath}
R.~P.~Feynman and A.~Hibbs, {\sl Quantum Mechanics and Path Integrals}
(McGraw Hill, New York, 1965).

\bibitem{lgt}
K.~G.~Wilson, {\sl Quarks And Strings On A Lattice},
in {\it New Phenomena In Subnuclear Physics}, Erice 1975 
(Plenum Press, New York, 1977).

\bibitem {parsour}
G.~Parisi and N.~Sourlas, {\it Phys.\ Rev.\ Lett.} {\bf 46} 871 (1981).

\bibitem {polymers}
T.~C.~Lubensky and J.~Isaacson, {\it Phys.\ Rev.} {\bf A20} 2130 (1979).
See also H.~Aoki, S.~Iso, H.~Kawai and Y.~Kitazawa,
{\it Prog.\ Theor.\ Phys.} {\bf 104} 877 (2000).

\bibitem {parisi_is}
H.~G.~Ballesteros, L.~A.~Fernandez, V.~Martin-Mayor, A.~Munoz Sudupe,
G.~Parisi and J.~J.~Ruiz-Lorenzo, hep-lat/9709078,
{\it Nucl.\ Phys.} {\bf B512} 681 (1998);
{\it J. Phys. Math. Gen.} {\bf A32} 1 (1999).

\bibitem {u1}
H.~W.~Hamber, {\it Phys.\ Rev.} {\bf D24} 941 (1981).

\bibitem {lang}
J.~Jersak, C.B.~Lang and T.~Neuhaus, 
{\it Phys.\ Rev.\ Lett.} {\bf 77} 1933 (1996).

\bibitem {lang1}
J.~Cox, W.~Franzki, J.~Jersak, C.B.~Lang, T.~Neuhaus and P.~Stephenson,
hep-lat/9608106 (1996).

\bibitem{schilling}
G.~Arnold, B.~Bunk, T.~Lippert and K.~Schilling,
{\it Nucl.\ Phys.\ Proc.\ Suppl.} {\bf B119} 864 (2003).

\bibitem{deser}
S.~Deser, R.~Jackiw and S.~Templeton,
{\it Annals Phys.} {\bf 140}, 372 (1982); {\it ibid.} {\bf 185}, 406 (1988); \\
S.~Deser, R.~Jackiw and G.~'t Hooft,
{\it Annals Phys.} {\bf 152}, 220 (1984).


\bibitem {corr}
H.~W.~Hamber, {\it Phys.\ Rev.} {\bf D50} 3932 (1994).

\bibitem {hdqg}
K.~S.~Stelle, {\it Phys. Rev.} {\bf D16} 953 (1977); \\
J.~Julve and M.~Tonin, {\it Nuovo Cimento} {\bf 46B} 137 (1978); \\
E.~S.~Fradkin and A.~A.~Tseytlin, {\it Phys.\ Lett.} 377 {\bf 104B} (1981)
and {\it ibid.} {\bf 106B} 63 (1981); {\it Nucl.\ Phys.} {\bf B201} 469 (1982); \\
I.~G.~Avramidy and A.~O.~Baravinsky, {\it Phys.\ Lett.} {\bf 159B} 269 (1985).

\bibitem {smit}
B.~de Bakker and J.~Smit, {\it Nucl.\ Phys.} {\bf B439} 239 (1995).

\bibitem{hasen}
A.~Hasenfratz and P.~Hasenfratz, {\it Phys.\ Lett.} {\bf B93} 165 (1980); \\
R.~Dashen and D.~Gross, {\it Phys.\ Rev.} {\bf D23} 2340 (1981). 

\bibitem {vipul}
V.~Periwal, Princeton preprint PUPT-1871, astro-ph/9906253 (June 1999);
see also T.~Goldman, J.~Perez-Mercader, F.~Cooper and M.~M.~Nieto,
{\it Phys.\ Lett.} {\bf B281} 219 (1992).



\end{thebibliography}
\end{document}